\documentclass[journal,draftclsnofoot,onecolumn]{IEEEtran}

\usepackage{setspace}
\usepackage{amsmath}
\usepackage{amssymb}
\usepackage{mathrsfs}
\usepackage{amsthm}
\usepackage{multirow}
\usepackage{color}
\usepackage{longtable}
\usepackage{array}
\usepackage{url}
\usepackage{comment}
\usepackage{enumerate}
\usepackage{eucal}
\usepackage{graphics}

\usepackage{algorithm}
\usepackage{algorithmicx}
\usepackage{algpseudocode}
\usepackage[noadjust]{cite}
\usepackage[table]{xcolor}

\usepackage{stmaryrd}

\usepackage{mathtools}

\DeclarePairedDelimiter\abs{\lvert}{\rvert}
\DeclarePairedDelimiter\ceil{\lceil}{\rceil}
\DeclarePairedDelimiter\floor{\lfloor}{\rfloor}
\DeclarePairedDelimiter\parenv{\lparen}{\rparen}

\DeclarePairedDelimiter{\set}{\{}{\}}

\newcommand{\cC}{\mathcal{C}}
\newcommand{\cD}{\mathcal{D}}
\newcommand{\cE}{\mathcal{E}}
\newcommand{\cF}{\mathcal{F}}

\newcommand{\cS}{\mathcal{S}}
\newcommand{\cT}{\mathcal{T}}

\newcommand{\cX}{\mathcal{X}}
\newcommand{\cY}{\mathcal{Y}}
\newcommand{\cZ}{\mathcal{Z}}

\newcommand{\ee}{\mathsf{e}}

\newcommand{\ban}{\bar{n}}

\renewcommand{\leq}{\leqslant}

\renewcommand{\geq}{\geqslant}

\newcommand{\ppmod}[1]{~({\rm mod~}#1)}

\newcommand{\Lm}{L_{\rm min}}
\newcommand{\Lo}{L_{\rm over}}

\theoremstyle{plain}
\newtheorem{theorem}{Theorem}
\newtheorem{corollary}[theorem]{Corollary}
\newtheorem{lemma}[theorem]{Lemma}
\newtheorem{proposition}[theorem]{Proposition}

\newtheorem{cons}{Construction}

\newenvironment{construction}{\begin{cons}}{\hfill$\openbox$ \end{cons}}

\theoremstyle{definition}
\newtheorem{definition}[theorem]{Definition}

\newtheorem*{remark}{Remark}

\newcommand{\N}{\mathbb{N}}

\newcommand{\vm}{\mathbf{m}}
\newcommand{\vp}{\mathbf{p}}

\newcommand{\vs}{\mathbf{s}}
\newcommand{\vv}{\mathbf{v}}
\newcommand{\vu}{\mathbf{u}}
\newcommand{\vw}{\mathbf{w}}
\newcommand{\vx}{\mathbf{x}}
\newcommand{\vy}{\mathbf{y}}

\newcommand{\vb}{\mathbf{b}}
\newcommand{\vc}{\mathbf{c}}

\newcommand{\baw}{\bar \vw}
\newcommand{\haw}{\hat \vw}
\newcommand{\bas}{\bar \vs}
\newcommand{\bam}{\bar \vm}
\newcommand{\hay}{\hat \vy}

\newcommand{\haS}{\hat{\mathcal{S}}}

\newcommand{\hcY}{\hat{\mathcal{Y}}}

\DeclareMathOperator{\wt}{wt}

\newcommand{\cCt}{\mathcal{C}_{\rm Trace}   }
\newcommand{\chCt}{\hat{\mathcal{C}}_{\rm Trace}   }
\newcommand{\pre}{\rm pre}
\newcommand{\suf}{\rm suf}

\newcommand{\eqdef}{\triangleq}

\DeclareMathOperator{\EncDist}{EncDist}

\begin{document}
\date{}
\title{Reconstruction from  Noisy Substrings}

\author{
  Hengjia Wei, Moshe Schwartz,~\IEEEmembership{Senior Member,~IEEE, and Gennian Ge}%
  \thanks{H. Wei is with the Peng Cheng Laboratory, Shenzhen 518055, China. He is also with the School of Mathematics and Statistics, Xi'an Jiaotong University, Xi'an 710049, China, and the Pazhou Laboratory (Huangpu), Guangzhou 510555, China
   (e-mail: hjwei05@gmail.com).}%
  \thanks{M. Schwartz is on a leave of absence from the School
    of Electrical and Computer Engineering, Ben-Gurion University of the Negev,
    Beer Sheva 8410501, Israel. He is now with the Department of Electrical and Computer Engineering at McMaster University, Hamilton, ON L8S 4K1, Canada
    (e-mail: schwartz.moshe@mcmaster.ca).}%
\thanks{G. Ge is with the School of Mathematical Sciences, Capital Normal University, Beijing 100048, China  (e-mail: gnge@zju.edu.cn).}
  \thanks{This work was supported in part by 
   the National Key Research and Development Program of China under Grant 2020YFA0712100, the National Natural Science Foundation of China under Grant 11971325, Grant 12231014 and Grant 12371523, Beijing Scholars Program, the major key project of Peng Cheng Laboratory under grant PCL2023AS1-2, and the Zhejiang Lab BioBit Program under Grant 2022YFB507.}
}

\maketitle

\begin{abstract}
This paper studies the problem of encoding messages into sequences which can be uniquely recovered from some noisy observations about their substrings. The observed reads comprise consecutive substrings with some given minimum
overlap. This coded reconstruction problem has applications to DNA storage. We consider both single-strand reconstruction codes and  multi-strand reconstruction codes, where the message is encoded into a single strand or a set of multiple strands, respectively. Various parameter regimes are studied.  New codes are constructed, some of whose rates asymptotically attain the upper bounds. 
\end{abstract}

\begin{IEEEkeywords}
    DNA storage, sequence (string) reconstruction, substitution, substring-distant sequences, robust positoining sequences.
\end{IEEEkeywords}

\section{Introduction} 
Sequence (string) reconstruction refers to a large class of problems of reconstructing a sequence from partial (perhaps noisy) observations of it. Instances of this problem include reconstruction from multiple erroneous copies of the sequence \cite{Levenshtein01TIT,Levenshtein01,BatKanKhaMcG:2004}, some substrings of the sequence \cite{KiaPulMil16,GabMil:2019}, all the length-$k$ subsequences \cite{ManMeySchSmiSto91,Sco97,DudSch03}, and compositions of the sequence's substrings or prefixes/suffixes \cite{AchDasMilOrlPan15,PatGabMil23}.

In this paper, we shall consider the  problem of encoding messages into sequences which can be uniquely recovered from observations about their substrings. This coding problem 
is motivated by  applications to DNA-based data storage systems, where data are encoded to long DNA sequences. In some DNA sequencing technologies (e.g., shotgun sequencing), a long DNA strand is first replicated multiple times, and these replicas are then fragmented into some short substrings  so that they could be  read. In order to retrieve the data, the original long sequence should be reconstructed based on the observations about these short substrings.

This coded reconstruction problem has been studied in different models with different assumptions on the substrings.  Gabrys and Milenkovic \cite{GabMil:2019} considered  the problem of reconstructing a  sequence of length $n$ from its \emph{$L$-multispectrum},  i.e., the multiset of all of its length-$L$ substrings. They constructed two classes of reconstruction codes with redundancies $2$ and $O(\log \log n)$ for $L>2 \log n$  and $\log n <L \leq 2 \log n$, respectively. They also studied the noisy settings in which some substrings/observations may be lost or be corrupted by errors, and constructed  codes to combat these effects.  Subsequently, Marcovich and Yaakobi \cite{MarYaa:2021} followed  this noisy setup  and provided more code constructions.  The constructions in \cite{GabMil:2019,MarYaa:2021} are based on the so-called  \emph{$(L,d)$-substring distant (SD) sequence}, a sequence in which  every two length-$L$ substrings are of Hamming distance at least $d$ apart. 
When $d=1$, such  sequences are also known as  \emph{$L$-substring unique  sequences} or  \emph{$L$-repeat free sequences}. Efficient encoding algorithms can be found in \cite{EliGabYaaMed:2021} for $L>\log n$. For general $d$, Marcovich and Yaakobi \cite{MarYaa:2021} proposed an encoding algorithm of $(L,d)$-SD sequences for $L>2\log n$.

Another   model is the \emph{torn-paper channel}, which  randomly tears the input sequence into small pieces of different sizes. 
The output of this channel is a set of substrings of  the input sequence with no overlap, and the message which is carried by the input sequence should be recovered  from these substrings.  This problem has been researched  in the
probabilistic setting in \cite{ShoVah:2021,RavVahSho:2021,NasShoVah:2022}. 
Recently, Bar-Lev \emph{et al.} \cite{BarMarYaaYeh:22} considered this problem in the worst-case. They studied both the noiseless setup and the noisy setup, and proposed a couple of index-based constructions to encode messages into sequences each of which can be uniquely recovered from its non-overlapping substrings. 
Furthermore, motivated by DNA sequencing technologies where multiple strings are sequenced simultaneously, they extended the single-strand reconstruction problem to a multi-strand reconstruction problem.  They constructed multi-strand reconstruction codes whose rates asymptotically behave like those  of single-strand reconstruction codes. Another related paper is by Wang \emph{et al.}~\cite{WanSimRav23}, which, unlike~\cite{BarMarYaaYeh:22}, does not restrict the length of the torn substrings, but rather their number. For this setting they construct codes that attain the upper bound on the rate up to asymptotically small factors.

In a recent paper, Yehezkeally \emph{et al.} \cite{YehBarMarYaa:2023} proposed a general model, which includes the two models above as extreme cases. In this model, the reconstruction is based on the sequence's \emph{$(\Lm,\Lo)$-trace}, which is  a multiset of subsrings where every substring has length at least  $\Lm$ and the overlap of every two consecutive substrings has length at least $\Lo$. They focused on the noiseless setup, and constructed a class of trace reconstruction codes whose rate can asymptotically achieve the upper bound. They also studied the multi-strand reconstruction problem in the $L$-multispectrum model, and proposed reconstruction codes whose rates are asymptotically $1$. 

In this paper, we shall follow the model in \cite{YehBarMarYaa:2023} and study the coding problem for both single-strand reconstruction and multi-strand reconstruction in the noisy setup. We aim to encode a message into a sequence which can be uniquely recovered  from its \emph{$(\Lm,\Lo,e)$-erroneous trace}, where each substring may suffer from at most $e$ substitution errors, or to encode a message into a set of $k$ sequences which can be recovered from the union of their $(\Lm,\Lo,e)$-erroneous traces.  Our contributions are listed as follows.

\begin{enumerate}
\item We first give an algorithm which can encode messages into $(L,d)$-SD sequences for $L=\ceil{a\log n}$ where $a>1$ is an arbitrary real constant. The rates of the encoded sequences asymptotically approach $1$. In contrast, the encoding algorithm  in \cite{MarYaa:2021}  requires  a single redundancy bit but works only when $L>2\log n$. 
\item For single-strand reconstruction, by using the proposed encoding algorithm for SD sequences, we construct two classes of $(\Lm,\Lo,e)$-trace reconstruction codes whose rates asymptotically achieve the upper bound.
\item For multi-strand reconstruction, we present some upper bounds on the rates of multi-strand $(\Lm,\Lo,e)$-trace reconstruction codes, as well as some code constructions. In some parameter regimes,  our constructions  yield codes whose rates asymptotically attain the upper bounds.
Interestingly, when $\log k=\kappa n$, $\Lm=a \log n$ and $\Lo=\gamma \Lm$, the maximal rates of multi-strand reconstruction codes  not only depend  on $\kappa,a,\gamma$, but also depend on the congruence class of $n$ modulo $\Lm-\Lo$.  
\end{enumerate}

\section{Preliminaries}
For a positive integer $n\in\N$, let $[n]$ denote the set $\set{0,1,2,\ldots, n-1}$. Let $\Sigma$ denote a finite alphabet. Throughout this paper, we always consider the binary case, i.e., $\Sigma=\set{0,1}$, however, our results can be easily generalized to non-binary cases. We use $\log x$ to denote the logarithm of $x$ to base $2$. When generalizing our results  to the $q$-ary alphabet case, it suffices to replace the $\log$ with $\log_q$. 

Assume $\vx=(x_0,x_1,\ldots,x_{n-1})\in \Sigma^n$ is a sequence over $\Sigma$. We denote its length $\abs{\vx}=n$, and its Hamming weight by $\wt_H(\vx)$. Given two sequence $\vx$ and $\vy$ over $\Sigma$, we denote their concatenation by  $\vx \circ \vy$. If $\vx$ and $\vy$ have the same length, we use $d_H(\vx,\vy)$ to denote their Hamming distance.

A \emph{substring} of $\vx$ is a sequence of the form $(x_a,x_{a+1},\dots,x_b)$, where $0\leq a\leq b<\abs{\vx}$, and we use $\vx[a,b]$ to denote it. We also use $\vx_{i+[L]}$, where $i\in[n-L+1]$, to denote the substring of $\vx$ which starts at the position $i$ and has length $L$, i.e., $\vx_{i+[L]}=(x_i,x_{i+1},\ldots,x_{i+L-1})=\vx[i,i+L-1]$.   

A code is simply a set $\cC\subseteq \Sigma^n$, whose elements are referred to as codewords. We say $n$ is the length of the code. The \emph{rate} of the code is defined as $R(\cC)=\frac{1}{n}\log \abs{\cC}$, and the redundancy of the code is $n-R(\cC)$.

\subsection{Reconstruction from the $L$-Multispectrum}
For a sequence $\vx \in \Sigma^n$ and a positive integer $L\leq n$, the \emph{$L$-multispectrum of $\vx$}, denoted by $\cS_L(\vx)$, is the \emph{multiset} of all its length-$L$ substrings, namely,
\[\cS_L(\vx) = \set*{\vx_{0+[L]},\vx_{1+[L]},\ldots,\vx_{n-L+[L]} }.\]

If $\vx$ can be uniquely reconstructed from its $L$-multispectrum, then we say it is \emph{$L$-reconstructible}. It was proved in \cite{Ukk:92} that if all the length-$(L-1)$ substrings of $\vx$ are distinct, then $\vx$ is $L$-reconstructible.  Such a sequence is referred to as an \emph{$L$-substring unique sequence}. In the works  \cite{GabMil:2019,EliGabYaaMed:2021}, algorithms were proposed   to construct a set of $L$-substring unique sequences of rate approaching $1$, where $L=\ceil{a \log n}$ for any constant real number $a >1$. 

In \cite{GabMil:2019}, Gabrys and Milenkovic further studied the problem of reconstructing sequences from their noisy multispectra.
They first considered the scenario where some substrings are not included in the readout spectrum.
For a subset $\haS\subset \cS_L(\vx)$, if the maximum number of consecutive substrings which are not included in $\haS$ is $G$, we say  $\haS$ has \emph{maximal coverage gap $G$}.
A code is called an \emph{$(L,G)$-reconstruction code} if every codeword $\vx$ can be uniquely reconstructed from any subset $\haS \subset \cS_L(\vx)$ with maximal coverage gap $G$. Gabrys and Milenkovic proposed a construction for such codes \cite{GabMil:2019} by restricting each codeword $\vx$ to be $\hat{L}$-substring unique with $\hat{L}<L-G$ and imposing some constraints on their prefixes.

 Gabrys and Milenkovic also researched the scenario where the observations about the  substrings suffer from substitution errors.   Let $\cY=\set{\vy_0,\vy_1,\ldots,\vy_{m-1}}$ be a multiset  consisting of $m$ strings of length $L$. If there is a subset $\haS=\set{\vx_{i_0},\vx_{i_1},\ldots,\vx_{i_{m-1}}}\subset\cS_L(\vx)$ with maximal coverage gap $G$ such that $d_H(\vy_j,\vx_{i_j})\leq e$ for all $j \in [m]$, then we say $\cY$ is an \emph{$(L,G,e)$-constrained erroneous multispectrum of $\vx$}. Moreover, $\cY$ is said to
be \emph{reliable} if for any symbol in $\vx$, there are more copies of the correct value rather than an incorrect value of the symbol. A code is called an \emph{$(L,G,e)$-reconstruction code} if  every codeword can be uniquely reconstructed from its any reliable $(L,G,e)$-constrained erroneous multispectrum\footnote{We emphasize
 that the multispectrum $\cY=\set{\vy_0,\vy_1,\ldots,\vy_{m-1}}$ is just a multiset, and the order/index $i$ of each $\vy_i$ cannot be directly read when reconstructing.}. Gabrys and Milenkovic constructed an $(L,G,e)$-reconstruction code of redundancy $O(\log \log n)$ for $L=6 \log n +O(\log \log n)$. Their construction is based on $(L,d)$-substring distant  sequences, whose definition is presented as follows.

\begin{definition}
\label{def:sd}
A sequence $\vw \in \Sigma^n$ is called  $(L,d)$-\emph{substring distant (SD)} if the minimum Hamming distance of its $L$-multispectrum is at least $d$, that is, $d_H(\vw_{i+[L]}, \vw_{j+[L]})\geq d$ for any $0\leq i <j \leq n-L$.
\end{definition}

\begin{remark}
We observe that an $(L,d)$-substring distant sequence is also $(L',d)$-substring distant, for any $L'\geq L$. Thus, we may equivalently say that $\vw \in \Sigma^n$ is $(L,d)$-\emph{substring distant (SD)} if $d_H(\vw_{i+[L']}, \vw_{j+[L']})\geq d$ for any integer $L'\geq L$ and $0\leq i <j \leq n-L'$. This equivalent definition allows $L$ to be a real number, which we shall conveniently use in the future.
\end{remark}

In \cite{MarYaa:2021}, Marcovich and Yaakobi followed the noisy setup of  Gabrys and Milenkovic. They studied the case of $G=0$, i.e., no substring losses.
Instead of reconstructing $\vx$ from a reliable erroneous multispectrum, they aimed to reconstruct from an $(L,0,e)$-erroneous multispectrum $\cY$,  the so-called \emph{maximum reconstructible-string}, i.e., a string of length $n$ that takes at every position $i$ the majority value of the occurrences of $x_i$ in $\cY$. Obviously, if $\cY$ is reliable, then the maximum reconstructible-string is equal to $\vx$. A sequence $\vx$ is called \emph{$(L,0,e)$-reconstructible}\footnote{The notion here is a bit different from that in \cite{MarYaa:2021}, where Marcovich and Yaakobi  further assumed that there are at most $t$ substrings in $\cY$ each of which is affected by at most $e$ errors and referred to it as a $(t,e)$-erroneous multispectrum.  They proposed two constructions for reconstructible codes:  one is independent of $t$ and thus can combat any number of erroneous substrings, while  the other one depends on $t$. In this paper, we focus on reconstructible codes which are independent of $t$.} if one can always reconstruct the maximum reconstructible-string from its any $(L,0,e)$-erroneous multispectrum.

\begin{proposition}[{{\cite[Theorem 16]{MarYaa:2021}}}]\label{prop:SDtorecon}
If $\vx$ is $(L-1,4e+1)$-SD, then it is $(L,0,e)$-reconstructible.
\end{proposition}

For positive integers $n,d,L$ with $d\leq L <n$, we use $\cZ_n(L,d)$ to denote the set of $(L,d)$-SD sequences of $\Sigma^n$. For fixed $d$ and $a >1$,  Marcovich and Yaakobi showed that the asymptotic rate of the set $\cZ(a \log n,d)$ is $1$, by using the Lov{\'a}sz Local Lemma.  Note that when $a<1$, even a single $(a \log n)$-substring unique sequence of length $n$ does not exist.

\begin{theorem}[{{\cite[Theorem 19]{MarYaa:2021}}}]
For fixed $d$ and $a>1$,
\[\lim_{n\to \infty} \frac{\log \abs{\cZ_n(a \log n,d)}}{n} =1.\]
\end{theorem}

Marcovich and Yaakobi also presented a deterministic algorithm which uses a single   redundancy bit to encode  $(a \log n,d)$-SD sequences for $a >2$.

\begin{theorem}[{{\cite[Algorithm~4 and Theorem~25]{MarYaa:2021}}}]Let $d>0$ be a fixed integer.   There is an encoding algorithm which uses a single redundancy bit to encode  $(L,d)$-SD sequences of length $n$, for \[L=2 \log n + 2(d-1+\epsilon) \log \log n,\]
where $\epsilon>0$ is a small constant number  and  $n$ is sufficiently large.
\end{theorem}

In Section~\ref{sec:SDseq}, we shall present an algorithm which can encode $(a \log n,d)$-SD sequences of length $n$ for any $a>1$, while its redundancy is $o(n)$. According to Proposition~\ref{prop:SDtorecon}, this implies  an $(L,0,e)$-reconstructible code whose rate approaches $1$, for $L=\ceil{a\log n}+1$ and $e =\floor{\frac{d-1}{4}}$.

\subsection{Reconstruction from an $(\Lm,\Lo)$-trace}

In \cite{YehBarMarYaa:2023}, Yehezkeally \emph{et al.} studied an extension of the problem of reconstructing from substrings.
Let $\vx\in \Sigma^n$ be a sequence.  A \emph{substring trace} of $\vx$ is a multiset of substrings $\set{\vx_{i_0+[L_0]}, \vx_{i_1+[L_1]},\ldots,\vx_{i_{m-1}+[L_{m-1}]}}$ for  some positive integer $m$, where $i_0 <i_1 <\cdots <i_{m-1}$. If $i_0=0$, $i_{j+1} <i_j+L_j$ for all $j<m-1$, and $i_{m-1}+L_{m-1}=n$, then the substring trace is called \emph{complete}.  Let $\Lm$ and $\Lo$ be two positive integers such that $\Lo < \Lm < n$. An \emph{$(\Lm,\Lo)$-trace} is a complete trace such that:
\begin{enumerate}
\item every substring has length at least $\Lm$, i.e., $L_i\geq\Lm$ for all $i\in[m]$;
\item the overlap of every two consecutive substrings has length at least $\Lo$, i.e., $i_j+L_j-i_{j+1} \geq \Lo$ for all $j\in[m-1]$.
\end{enumerate}

For a sequence $\vx$, let $\cT_{\Lm}^{\Lo}(\vx)$ denote the set of all $(\Lm,\Lo)$-traces of $\vx$. A code $\cC$ is referred to as an \emph{$(\Lm,\Lo)$-trace reconstruction code} if $\cT_{\Lm}^{\Lo}(\vx)\cap \cT_{\Lm}^{\Lo}(\vx')=\emptyset$ for all $\vx\neq \vx'\in \cC$, or equivalently, every codeword can be uniquely reconstructed from any of its $(\Lm,\Lo)$-traces.

\begin{proposition}[{{\cite[Lemma 1]{YehBarMarYaa:2023}}}]\label{prop:UStorecon}
Let $\vx$ be an $\Lo$-substring unique sequence. Then $\vx$ can be uniquely reconstructed from any of its $(\Lm,\Lo)$-traces.
\end{proposition}

By refining the constructions of substring unique sequences, Yehezkeally \emph{et al.} obtained the following result.

\begin{theorem}[{{\cite[Corollary 6]{YehBarMarYaa:2023}}}]\label{thm:tracereconerrfree-1}
There is an $(\Lm,\Lo)$-trace reconstruction code of $\Sigma^n$ whose rate approaches $1$, for $\Lo \geq \ceil{\log n} + 3\ceil{ \log \log n} +12$ and sufficiently large $n$.
\end{theorem}

They also studied the other parameter regimes.

\begin{lemma}[{{\cite[Lemma 8]{YehBarMarYaa:2023}}}]\label{lm:upbndsubstringprev}
If $\Lm=a\log n +O(1)$ and $\Lo=\gamma \Lm+O(1)$ for some $a>1$ and $0\leq \gamma \leq \frac{1}{a}$, then for any $(\Lm,\Lo)$-trace reconstruction code $\cC\subseteq \Sigma^n$, its rate $R(\cC)$ must satisfy
\[R(\cC) \leq \frac{1-1/a}{1-\gamma} +O\parenv*{\frac{\log \log n }{ \log n} }. \]
\end{lemma}

\begin{theorem}[{{\cite[Theorem 15]{YehBarMarYaa:2023}}}]\label{thm:tracereconerrfree-2} Let $\Lm=a\log n$ and $\Lo=\gamma \Lm$ for some $a>1$ and $0\leq \gamma \leq \frac{1}{a}$.  If $n$ is sufficiently large,   then there is an  $(\Lm,\Lo)$-trace reconstruction code $\cC\subseteq \Sigma^n$ with rate
\[R(\cC) \geq \frac{1-1/a}{1-\gamma} - \frac{(\log n)^\epsilon }{ a \sqrt{\log n}} - O\parenv*{\frac{1}{\sqrt{\log n}}}, \]
where $\epsilon>0$ is a small number which is independent of $n$.
\end{theorem}

In this paper, we shall study the problem of  reconstructing sequences  from their noisy substring traces.
Let  $\cY=\set{\vy_0,\vy_1,\ldots,\vy_{m-1}}$ be a multiset of sequences  over $\Sigma$, and let $L_j=\abs{\vy_j}$ for $j \in [m]$. We say $\cY$ is an \emph{$(\Lm,\Lo,e)$-erroneous trace of $\vx$} if there exists an $(\Lm,\Lo)$-trace $\set{\vx_{i_0+[L_0]}, \vx_{i_1+[L_1]},\ldots,\vx_{i_{m-1}+[L_{m-1}]}}$ such that  $d_H(\vy_{j},\vx_{{i_j}+[L_j]} )\leq e$ for all $j \in [m]$. Namely, each string $\vy_j$ in $\cY$ is an erroneous copy of the substring $\vx_{{i_j}+[L_j]}$ in $\vx$ with at most $e$ errors. The index $i_j$ is referred to as the \emph{location $\vy_j$ in $\vx$}. For a sequence $\vx$ and its any $(\Lm,\Lo,e)$-erroneous trace $\cY$, if one can always determine the location of every $\vy_i \in \cY$ in $\vx$, then we say $\vx$ is \emph{$(\Lm,\Lo,e)$-trace reconstructible}.  We note that  once all the locations of $\vy_j$'s are identified, the maximum reconstructible-string of $\cY$ can be determined by taking  at every position $i$ the majority value of the occurrences of $x_i$ in $\cY$. Hence, the $(\Lm,\Lo,e)$-trace reconstructible sequence $\vx$ can be uniquely reconstructed as long as $\cY$ is reliable.

A code is  called an \emph{$(\Lm,\Lo,e)$-trace reconstruction code} if  every codeword $\vx$ is $(\Lm,\Lo,e)$-trace reconstructible\footnote{Unlike the noiseless case, in  an $(\Lm,\Lo,e)$-trace reconstruction code it might be possible that two codewords share a common $(\Lm,\Lo,e)$-erroneous trace. Nevertheless,  they cannot have a  common reliable trace. }.
In Section~\ref{Sec:tracecode}, we will give two constructions for $(\Lm,\Lo,e)$-trace reconstruction codes where the number of errors $e$ is fixed. Our results are akin to Theorem~\ref{thm:tracereconerrfree-1} and Theorem~\ref{thm:tracereconerrfree-2}. In particular, when $\Lo=a\log n$ for some $a>1$, we construct a class of $(\Lm,\Lo,e)$-trace reconstruction codes whose rates approach $1$. When $\Lm= a\log n$ and $\Lo=\gamma \Lm$  for some $a>1$ and $0\leq \gamma \leq \frac{1}{a}$, the proposed $(\Lm,\Lo,e)$-trace reconstruction codes have rates close to $\frac{1-1/a}{1-\gamma}$. These  results are summarized in Table~\ref{tab:summary-3}.
Our constructions are based on robust positioning sequences and window-weight limited sequences, which are reviewed in Section~\ref{subsec:RPS}.

We note that when $\Lo=0$, $(\Lm,0)$-reconstruction codes were researched by Bar-Lev \emph{et al.} in \cite{BarMarYaaYeh:22} by the name of \emph{adversarial torn-paper codes}. In the same paper, they also consider the scenario where the DNA strand may suffer from  substitution errors \emph{before} sequencing.  Such kind of errors cannot be corrected by majority decoding. Yehezkeally and Polyanskii studied a similar problem for the $(L+1,L)$-trace reconstruction \cite{YehPol23}. They introduced the notion of \emph{$(t,L)$-resilient repeat free sequence}, which satisfies the property that the result of any $t$ substitution errors to it is $L$-repeat free, and  proposed an algorithm to directly encode such sequences. Interestingly,  \cite[Lemma 6]{YehPol23} shows that an $(L,2t+1)$-SD sequence is $(t,L)$-resilient repeat free. In Section~\ref{Sec:tracecode},  we will also study errors before sequencing and  modify our code construction for $(\Lm,\Lo,e)$-trace reconstruction to combat  such errors.

\begin{table*}
 \caption{Lower and upper bounds on the code rate of  single-strand $(\Lm,\Lo,e)$-trace reconstruction codes of $\Sigma^n$.}
  \label{tab:summary-3}
{\small
  {\renewcommand{\arraystretch}{1.5}
  \begin{tabular}{cccccc}
    \hline\hline
    Parameter regimes   & Lower bound & Ref. & Upper bound & Ref.   \\
    \hline
     $\Lo= \ceil{\log n} +(6d+7) \ceil{\log \ceil{\log n}}+ d \lceil \log d \rceil+5d$   & \multirow{2}*{ $1-o(1)$} &  \multirow{2}*{Corollary~\ref{cor:genrecon-1}} & \multirow{2}*{ $1$ }&   \\
     where $d=4e+1$ & & & & \\
    \hline
     $\Lm= \ceil{a \log (n)}$, $\Lo= \ceil{\gamma \Lm}$ &  \multirow{2}*{$\frac{1-1/a}{1-\gamma} -o(1)$} &  \multirow{2}*{Theorem~\ref{thm:tracerecon}} &  \multirow{2}*{$\frac{1-1/a}{1-\gamma} +o(1)$} &  \multirow{2}*{Lemma~\ref{lm:upbndsubstringprev}} \\
     where $a>1$ and $0\leq a\gamma \leq 1$  &  & \& Theorem~\ref{thm:gamma0} &  & \\ 
    \hline\hline
  \end{tabular}
  }
  }
\end{table*}

\subsection{Multi-strand reconstruction} 
Motivated by DNA sequencing technologies where multiple DNA strands are sequenced simultaneously, 
the reconstruction problem has been extended to the multi-strand case in \cite{YehBarMarYaa:2023,BarMarYaaYeh:22}, i.e., reconstructing a \emph{multiset} of $k$ sequences of length $n$ from the union of their traces.

Define 
\[\cX_{n,k} \eqdef \set*{  \set*{\vx_0,\vx_1,\ldots,\vx_{k-1}} ~:~ \vx_i \in \Sigma^{n} \textup{ for all } i \in [k] }.\]
 Then $\abs{\cX_{n,k}} = \binom{k +2^n-1}{k}$. The \emph{rate} of a multi-strand code $\cC\subseteq \cX_{n,k}$  is defined as 
\[R(\cC)\eqdef \frac{\log \abs{\cC}}{ \log \abs{\cX_{n,k}} }. \]

For a multiset $\cS=\set{\vx_0,\vx_1,\ldots,\vx_{k-1}}\in \cX_{n,k}$, its \emph{$(\Lm,\Lo)$-trace} is a (multiset) union  $\cY=\bigcup_{i=0}^{k-1}\cY_i$, where each $\cY_i$ is an $(\Lm,\Lo)$-trace of $\vx_i$. A code $\cC\subseteq \cX_{n,k}$ is referred to as a \emph{multi-strand $(\Lm,\Lo)$-trace reconstruction code} if every codeword can be reconstructed from its $(\Lm,\Lo)$-trace. 
Two classes of multi-strand trace reconstruction codes whose rates asymptotically attain the upper bound have been constructed in \cite{YehBarMarYaa:2023,BarMarYaaYeh:22}, for $\Lo=0$ or $\Lo=\Lm-1$, respectively. 

\begin{theorem}[{{\cite[Theorem 12]{BarMarYaaYeh:22}}}]
Suppose that $\log k =o(n)$ and $\Lm=a \log (nk)$ with $a>1$. Then there is a class of multi-strand $(\Lm,0)$-trace reconstruction codes of rate $1-1/a-o(1)$.
\end{theorem}

\begin{theorem}[{{\cite[Corollary 23]{YehBarMarYaa:2023}}}]
Suppose that  $\limsup_{n\to \infty} \log k / n <1$ and $\Lm\geq \log(nk) +3\log \log (nk)+12$. Then there is a class of multi-strand $(\Lm,\Lm-1)$-trace reconstruction codes of  rate $1-o(1)$.
\end{theorem}


In this paper, we will also study the problem of reconstructing multiple strands from their noisy traces. 
For a multiset $\cS=\set{\vx_0,\vx_1,\ldots,\vx_{k-1}}\in \cX_{n,k}$, its \emph{$(\Lm,\Lo,e)$-erroneous trace} is a (multiset) union  $\cY=\bigcup_{i=0}^{k-1}\cY_i$, where each $\cY_i$ is an $(\Lm,\Lo,e)$-erroneous trace of $\vx_i$. We aim to reconstruct $\cS$ from its  $(\Lm,\Lo,e)$-erroneous trace. If for  any   $(\Lm,\Lo,e)$-erroneous trace $\cY$ of $\cS$ and any $\vy\in\cY$, it is possible to determine the index $i$ such that $\vy\in\cY_i$ as well as the location of $\vy$ in $\vx_i$, then we say $\cS$ is \emph{$(\Lm,\Lo,e)$-trace reconstructible.} 
A code $\cC\subseteq \cX_{n,k}$ is  called an \emph{multi-strand $(\Lm,\Lo,e)$-trace reconstruction code} if each of its codewords is $(\Lm,\Lo,e)$-trace reconstructible.

Following the research in \cite{YehBarMarYaa:2023}, we assume that  $\limsup_{n\to \infty} \log k / n <1$, which is of great interest in applications. In Section~\ref{sec:mulstrrecon}, we shall present some upper bounds on the multi-strand trace code rate and propose some codes whose rates asymptotically attain these bounds. Our results are summarized in Table~\ref{tab:summary-1} and Table~\ref{tab:summary-2}. Among others, when $\log k =\kappa n$ with $0<\kappa <1$, we obtain  a class of multi-strand $(\Lm,0,e)$-trace reconstruction codes of rate $\frac{1-1/a}{ 1-\kappa}+ \frac{L^*}{a(1-\kappa)n} -o(1)$, where  $L^*\equiv n \pmod{\Lm}$. Note that $L^*\in [\Lm]$ and $\Lm=a \log (nk)=\Theta(n)$. The term $\frac{L^*}{n}$ could be a non-vanishing number, depending on the congruence class of $n$ modulo $\Lm$. 
In contrast, when $\log k =o(n)$, the rate of the multi-strand $(\Lm,0)$-trace reconstruction  codes in  \cite[Theorem 12]{BarMarYaaYeh:22} is $1-1/a-o(1)$, which is the same as that of single-strand reconstruction codes.

\begin{table*}
 \caption{Lower and upper bounds on the code rate of  multi-strand $(\Lm,\Lo,e)$-trace reconstruction codes of $\cX_{n,k}$, where $\log k =o(n)$.}
  \label{tab:summary-1}
{\small
  {\renewcommand{\arraystretch}{1.5}
  \begin{tabular}{cccccc}
    \hline\hline
    Parameter regimes   & Lower bound & Ref. & Upper bound & Ref.   \\
    \hline
     $\Lo= \log (nk) +(24e+13)\log \log (nk)+O(1)$   & $1-o(1)$ &  Theorem~\ref{thm:multilargeLo} & $1$ &   \\
    \hline
     $\Lm=\ceil{a \log (nk)}$, $\Lo=\ceil{\gamma \log (nk)}$ &  \multirow{2}*{$\frac{1-1/a}{1-\gamma} -o(1)$} &  \multirow{2}*{Theorem~\ref{thm:multismallLoLowBnd}} &  \multirow{2}*{$\frac{1-1/a}{1-\gamma} +o(1)$} &  \multirow{2}*{Lemma~\ref{lemma:multismallLoUpBnd}} \\
     where $a>1$ and $0\leq a\gamma \leq 1$  &  &  &  & \\ 
     \hline
     $\Lm\leq \log (nk)+o(\log (nk))$ &  & & $o(1)$ &  Corollary~\ref{cor:vanishinglowbnd-1}\\
    \hline\hline
  \end{tabular}
  }
  }
\end{table*}

\begin{table*}
 \caption{Lower and upper bounds on the code rate of multi-strand $(\Lm,\Lo,e)$-trace reconstruction codes of $\cX_{n,k}$, where $\log k =\kappa n$ and $L^* = (n-\Lo)\bmod (\Lm-\Lo)$}
  \label{tab:summary-2}
{\small
  {\renewcommand{\arraystretch}{1.5}
  \begin{tabular}{cccccc}
    \hline\hline
    Parameter regimes   & Lower bound & Ref. & Upper bound & Ref.   \\
    \hline
     $\Lo= \log (nk) +(24e+13)\log \log (nk)+O(1)$   & $1-o(1)$ &  Theorem~\ref{thm:multilargeLo} & $1$ &   \\
    \hline
     $\Lm=\ceil{a \log (nk)}$, $\Lo=\ceil{\gamma \Lm}$ &  \multirow{2}*{$\frac{1-a\gamma \kappa}{ 1-\kappa} \parenv*{\frac{1-1/a}{1-\gamma}} -o(1)$} &  \multirow{2}*{Theorem~\ref{thm:multismallLoLowBnd}} &  {$\frac{1-a\gamma \kappa}{ 1-\kappa} \parenv*{\frac{1-1/a}{1-\gamma}}$} &  \multirow{2}*{Lemma~\ref{lemma:multismallLoUpBnd}} \\
     where $a>1$ and $0\leq a\gamma \leq 1$  &  &  & $+\frac{1/a-\gamma}{(1-\gamma)(1-\kappa)} \frac{L^*}{n} +o(1)$ & \\
     \hline
      $\Lo=0$, $\Lm=\ceil{a \log (nk)}$, $a>1$,  &  \multirow{2}*{$\frac{1-1/a}{ 1-\kappa}+ \frac{L^*}{a(1-\kappa)n} -o(1)$} &  \multirow{2}*{Theorem~\ref{thm:multinullLoLowBnd}} &  \multirow{2}*{$\frac{1-1/a}{ 1-\kappa}+ \frac{L^*}{a(1-\kappa)n} +o(1)$} &  \multirow{2}*{Lemma~\ref{lemma:multismallLoUpBnd}} \\
      and $L^* \leq \Lm-(1+\epsilon)\log(nk)$  &   &  &  & \\
      \hline 
     $\Lm = \log (nk)+ o(\log (nk))$   &  & & \multirow{2}*{$o(1)$} & \multirow{2}*{Lemma~\ref{lm:vanishinglowbnd-3}} \\
     and $\Lm-\Lo=\Theta(\log(nk))$\\
     \hline
     $\Lm = \ceil{a\log (nk)}$ with $a<1$ &  & & $o(1)$ &  Lemma~\ref{lm:vanishinglowbnd-2}\\
    \hline\hline
  \end{tabular}
  }
  }
\end{table*}

\subsection{Robust positioning sequences}\label{subsec:RPS}

An $(L,d)$-substring distant  sequence $\vx$ is also known as an \emph{$(L,d)$-robust positioning sequence}, since  the contents of any length-$L$ substring  can locate the substring's position in $\vx$, even if they are corrupted by at most $\floor{(d-1)/2}$ errors. In the context of robust positioning sequences,  given $L$ and $d$, it is of interest to construct a (single) long $(L,d)$-robust positioning sequence with  efficient locating algorithm.
This problem, as well as its 2-dimensional extension, has been discussed in \cite{bruckstein2012simple,BerkowitzKopparty:2016,Cheeetal:2020,Daoetal2020,Wei:2022}. Among others, Chee \emph{et al.} \cite{Cheeetal:2020} constructed a class of $(L,d)$-robust positioning sequences of length ${2^L}/(c L^{3d+6.5})$ for some constant number $c>0$. Their construction was refined in \cite{Wei:2022} to obtain  sequences of length ${2^L}/(c L^{\ceil{(d-1)/2}+8})$, which is nearly optimal.   The constructions in \cite{Cheeetal:2020,Wei:2022} require the following notions.

\begin{theorem}[$d$-Auto-Cyclic Sequences \cite{LevYaa:2017}]\label{thm:acv}
Let  $\ell=d \lceil \log d \rceil+2d$. Set $\vu$ to be the sequence
\begin{equation*}
 \vu  =1^d \circ \vu_0 \circ \vu_1 \circ \cdots  \circ \vu_{\lceil \log d \rceil}, \mbox{ where }  \vu_i  =((1^{2^i}\circ0^{2^i})^d)[0,d-1].
\end{equation*}
Then for all $1\leq i \leq d$, we have that  $$d_H(\vu,0^i\circ \vu[0,{\ell-i-1}])\geq d,$$ and $\vu$ is called a {\it $d$-auto-cyclic sequence}.
\end{theorem}

\begin{definition}
Let $n,L,d$ be positive integers such that $d<L<n$. We say a sequence $\vx\in \Sigma^n$ satisfies the $(L,d)$-\emph{window weight limited (WWL)} constraint, and is called an \emph{$(L,d)$-WWL sequence},
if  $\wt_H(\vx_{i+[L]})\geq d$ for any $i\in[n-L+1]$.
\end{definition}

\begin{proposition}[{{\cite[Construction~1 and Theorem~3.7]{Cheeetal:2020}}}]\label{prop:conSDS}
Given $L$ and $d$, choose $K$ such that $\ell<K$ and $K+\ell<L$, where $\ell=d \lceil \log d \rceil+2d$.
Let $\vu$ be a $d$-auto-cyclic vector of length $\ell$  from Theorem~\ref{thm:acv} and set
$L_p=K+\ell$. 
Let $\vs_0,\vs_1,\ldots,\vs_{M-1}$ be a collection of length-$(L-L_p)$  binary vectors satisfying the following conditions:
\begin{enumerate}
\item[(P1)] $\vs_i$ is a $(K,d)$-WWL vector for $i\in [M]$;
\item[(P2)] $\vs_{i+1}[0,j-1]\circ \vs_i[j,L-L_p-1]$ is a $(K,d)$-WWL vector for $i\in[M-1]$ and $j\in [L-L_p-1]$; and
\item[(P3)] the concatenation $\vs_0\circ\vs_1\circ\vs_2\circ\cdots\circ\vs_{M-1}$ is an $(L-L_p,d)$-modular robust positioning sequence\footnote{A sequence $\vw$ is an $(L-L_p,d)$-modular robust positioning sequence if $d_H(\vw_{i+[L-L_p]},\vw_{j+[L-L_p]})\geq d$ for any $i\equiv j \pmod{L-L_p}$ and $i\neq j$.}.
\end{enumerate}
Then the sequence
\[\vs\triangleq 0^K\circ \vu \circ \vs_0 \circ 0^K\circ \vu \circ \vs_1  \circ \cdots \circ 0^K \circ \vu \circ \vs_{M-1}\]
is an $(L,d)$-robust positioning (substring distant) sequence.
\end{proposition}

\begin{theorem}[{{\cite[Construction~1A and Corollary~3.12]{Cheeetal:2020}}}]\label{thm:singleSDS}
Given $d$ and $L$, set  $K=3\ceil{(3\log L)/2}=\frac{9}{2}\log L+O(1)$.   There is an explicit construction   of  sequences $\vs_0,\vs_1,\ldots,\vs_{M-1}$ of length $L-K-\ell$, where  $\log M =L-3d\log L -7.5 \log L -O(1)$,  such that the conditions (P1)--(P3) in Proposition~\ref{prop:conSDS} are satisfied. 
\end{theorem}

\begin{remark}
We note  that  for each $i \in [M]$, the concatenation $0^K\circ \vu \circ \vs_i$ is an $(L_p,d)$-WWL sequence, since the length-$d$ prefix of $\vu$ is $1^d$ and $\vs_i$ is $(K,d)$-WWL.
\end{remark}

\section{Encoding of $(a \log n,d)$-Substring Distant Sequences for $a>1$}\label{sec:SDseq}

In this section we shall present an encoding method which can generate a set of $(a\log n,d)$-SD sequences of length $n$ (with $a>1$, a real number)  whose rate asymptotically approaches $1$. We shall, in fact, construct $(L,d)$-SD sequences with $L=\log n + (6d+7)\log \log n +O(1)$, but using the remark following Definition~\ref{def:sd}, we shall find it more convenient to denote these sequences as $(a\log n,d)$-SD.

We first require some notations. For a sequence $\vw \in \Sigma^n$, we say that $(i,j)$ (where $0\leq i<j\leq n-L$) is an $(L,\rho)$-\emph{close window pair} in $\vw$ if  $d_H(\vw_{i+[L]},\vw_{j+[L]})\leq \rho$. Moreover,  $(i,j)$ is called \emph{primal}, if for any other $(L,\rho)$-close window pair $(i',j')$ in $\vw$ we have $j\leq j'$.
Let $\vx,\vx' \in \Sigma^L$ be two sequences with $d_H(\vx,\vx')\leq \rho$ for some integer $\rho \leq L$. Let $p_{1}, p_{2},\ldots,p_{d_H(\vx,\vx')} $ denote the indices of the entries where $\vx$ and $\vx'$ do not agree.
For every ${1\leq i\leq \rho}$ let
\begin{equation}\label{eq:linlocalenc}
  \vb_i=
  \begin{cases}
    b(p_i) & \text{if $i\leq d_H(\vx,\vx')$}, \\
    0^{{\ceil{\log (L+1)}}} & \text{otherwise},
  \end{cases}
\end{equation}
where $b(i)$ is  the  binary representation of $i$ with ${\ceil{\log (L+1)}}$ symbols.
Let
\[\EncDist_{L,\rho}(\vx,\vx') \eqdef \vb_1 \circ \vb_2 \circ \cdots \circ \vb_{\rho}.\]
Then $\EncDist_{n,\rho}(\vx,\vx')$ encodes the difference between $\vx$ and $\vx'$, and  its length  is $\rho {\ceil{\log (L+1)}}$.

Given a fixed $d$ and a sufficiently large $n$, we are going to present an encoding algorithm which can encode   $(L,d)$-SD sequences of length $n$. Set 
\begin{align*}
L_1&\eqdef \ceil{\log n} +(2d-1)\ceil{ \log \ceil{\log n}} +6d+\ceil{\log (d+1)},\\
K_1&\eqdef d\ceil{\log \ceil{\log n}}+d,\\
L_2&\eqdef \ceil{\log n}+(3d+7)\ceil{\log \ceil{\log n}},\\
K_2&\eqdef 3\ceil*{\frac{3}{2}\log L_2},\\
K_{\max} & \eqdef \max\set{K_1,K_2}.
\end{align*}
Additionally, set 
\begin{align*}
\ell&\eqdef d \lceil \log d \rceil+2d,\\
L&\eqdef \max\{L_1+K_2+K_{\max}+\ell, L_2+2K_1+K_{\max}+\ell\}.
\end{align*}
Assume that  $d$ is fixed and $n$ is sufficiently large. Then $L=L_2+2K_1+K_{\max}+\ell$, and $K_1 > K_2$ if and only if $d\geq 5$. Note that
\[K_2=3 \ceil*{1.5\log L_2 } 
\leq 3\ceil*{ 1.5\log \ceil*{\log n} +1.5 }\leq 4.5 \ceil*{\log \ceil*{\log n}} +7.5.\]
Thus, we have that 
\[
L \begin{cases}
   =\ceil{\log n} +(6d+7) \ceil{\log \ceil{\log n}}+ d \lceil \log d \rceil+5d & \text{if $d\geq 5$}, \\
   \leq \ceil{\log n} +(5d+11.5) \ceil{\log \ceil{\log n}}+ d \lceil \log d \rceil+4d+7.5  & \text{otherwise.}
  \end{cases}
\]
Our encoder   resembles the encoding algorithms in \cite{GabMil:2019,EliGabYaaMed:2021} and consists of the following three parts:
\begin{enumerate}
\item We first  use the encoder presented in \cite{LevYaa:2017} to encode a message sequence $\vm \in \Sigma^{n'}$ into a $(d\ceil{\log \ceil{\log (n)}},d)$-WWL sequence $\vw$ of length $n-K_1-K_2$. According to \cite[Corollary 20]{LevYaa:2017}, this encoder, denoted by $\cE_1$, requires approximately $2d \cdot 2 ^{\cF(n-K_1-K_2,d)- d\ceil{\log \ceil{\log n}}}$ redundancy symbols, where \[\cF(n,d)=\ceil{\log n}+(d-1) (\ceil{\log \ceil{\log n}}+C)+2 \]
for some constant $C$.
Hence, 
\begin{equation} \label{eq:redundancy}
n' = n-K_1-K_2- 
 2d \cdot 2 ^{\cF(n-K_1-K_2,d)- d\ceil{\log \ceil{\log n}}}= n-K_1-K_2-\Theta(n/\log n).
 \end{equation}
\item Then we encode the $(d\ceil{\log \ceil{\log n}},d)$-WWL sequence $\vw$ into an $(L_1,d)$-SD sequence $\baw$ by eliminating the pairs of substrings of small distance and attaching some information about their positions and difference.  This encoder, denoted by $\cE_2$, is presented in Algorithm~\ref{alg:Elim}, and it can   additionally guarantee the output sequence is $(K_1,d)$-WWL.
\item   As an output of Algorithm~\ref{alg:Elim}, the sequence $\baw$ is usually shorter than the sequence $\vw$. Thus, we need an  expansion step to increase the sequence length while keeping the substring-distant property. Let  $\vs_0,\vs_1,\ldots,\vs_{M-1}$ be a collection of $(K_2,d)$-WWL sequences of length $L_2-L_p$ as in Theorem~\ref{thm:singleSDS}. Set
\[\bas \eqdef 0^{K_{\max}} \circ \vu \circ \vs_0 \circ 0^{K_{\max}} \circ \vu \circ\vs_1 \circ \cdots \circ 0^{K_{\max}} \circ \vu \circ \vs_{M-1},\]
where $\vu$ is the $d$-auto-cyclic vector of length $\ell$ from Theorem~\ref{thm:acv}.
Finally, let \[ \haw  \eqdef \cE_3(\baw)\eqdef (\baw \circ 0^{K_2} \circ \bas)[0,n-1].\]
\end{enumerate}
We shall show  $\haw$ is the required  $(L,d)$-SD sequence of length $n$.

We first describe the encoding  presented in Algorithm~\ref{alg:Elim}. This procedure  encodes a $(d\ceil{\log \ceil{\log n}}, d)$-WWL sequence $\vw$ into a sequence $\baw$ that is simultaneously $(L_1,d)$-SD and $(K_1,d)$-WWL.  Initiate $\baw=\vw$. If there are no $(L_1,d-1)$-close window pairs in $\baw$, then the algorithm returns $\baw$ as the output. We observe that since $\vw$ is $(d\ceil{\log \ceil{\log n}},d)$-WWL and $K_1\geq d\ceil{\log \ceil{\log n}}$, then $\vw$ is also $(K_1,d)$-WWL.

Otherwise, we choose a  primal $(L_1,d-1)$-close window pair, say $(i,j)$. We replace the substring $\baw_{j+[L_1]}$ with the sequence
\begin{equation}\label{eq:insertedseq-1}
1^d \circ 0^{d\ceil{\log \ceil{\log n}}} \circ 1^d \circ B(i) \circ 1^d \circ \EncDist_{L_1,d-1}(\baw_{i+[L_1]}, \baw_{j+[L_1]})   \circ 0^{\ceil{\log (d+1)}} \circ 1^d,
\end{equation}
where $B(i):[n] \longrightarrow \Sigma^{\ceil{\log n} +d}$ is the encoding function in \cite[Algorithm 2]{LevYaa:2017}, which can encode integers in $[n]$ into $(d\ceil{\log \ceil{\log n}},d)$-WWL sequences  in $O(n)$ time. We note that this sequence is $(K_1,d)$-WWL and contains the information about the position $i$ and the difference between $\baw_{i+[L_1]}$ and $\baw_{j+[L_1]}$. Moreover, the substring $0^{d\ceil{\log \ceil{\log n}}}$ serves as a marker which indicates the position $j$ of the removed substring $\baw_{j+[L_1]}$.

We shall repeat this procedure until there are no   $(L_1,d-1)$-close window pairs in $\baw$. But in order to  ensure that $\vw$ can be recovered from the output of the algorithm, we  need more tricks. We note that in \cite{GabMil:2019}  the inserted sequences always start with a marker $0^{2\log \log n}$ and end with a symbol `$1$'. This  pattern together with the rule that only the primal pairs can be chosen and replaced guarantees  that after each replacement  the latest inserted substring always starts with the rightmost $0^{2\log \log n}$ in $\baw$. Due to this property, we have a decoding algorithm which can recover $\vw$ from $\baw$: Let $\baw^{(k)}$ denote the sequence $\baw$ after the $k$-th replacement. One can search for the rightmost $0^{2\log \log n}$ in $\baw^{(k)}$ to find the position $j$ of the inserted substring in the $k$-th replacement. By replacing the inserted substring with the removed substring, one can recover $\baw^{(k-1)}$ from $\baw^{(k)}$.  Doing this iteratively,  one can eventually recover $\vw$ from $\baw$.

 In our encoding, the inserted substring should always  contain  $1^d$ as both prefix and suffix to maintain the property of being $(K_1,d)$-WWL. We have to modify the substring $0^{\ceil{\log (d+1)}}$ in \eqref{eq:insertedseq-1} to ensure  the latest inserted substring always starts with the rightmost $1^d\circ 0^{d\ceil{\log\ceil{ \log n}}}$ in $\baw$.
 Let $j_p$ and $j$ be the positions of the removed substrings in the previous replacement and in the current replacement, respectively. Since we only choose the primal pairs, necessarily,  $j>j_p-L_1$. If $j>  j_{p}-L_1+d$, then we still replace the substring $\baw_{j+[L_1]}$ with the sequence in \eqref{eq:insertedseq-1}, since the  marker $0^{d \ceil{\log \ceil{\log n}}}$  which is inserted in the previous replacement will be destroyed by the suffix $1^d$ of this inserted sequence. If  $j_p-L_1 < j \leq  j_{p}-L_1+d$, we first set $\baw[j_p+d]$ to be `1' to destroy the previous marker $0^{d\ceil{\log \ceil{\log n}}}$. Then we replace  $\baw_{j+[L_1]}$ with the sequence
\begin{equation}\label{eq:insertedseq-2}
1^d \circ 0^{d\ceil{\log \ceil{\log n}}} \circ 1^d \circ B(i) \circ 1^d \circ \EncDist_{L_1,d-1}(\baw_{i+[L_1]}, \baw_{j+[L_1]}) \circ b(j-j_p+L_1) \circ 1^d,
\end{equation}
where $b(j-j_p+L_1)$ is the binary encoding of $j-j_p+L_1$ with $\ceil{\log (d+1)}$ symbols, since $1\leq j-j_p+L_1\leq d$.

Note that the substring $B(i)$ and  the substring $\EncDist_{L_1,d-1}(\baw_{i,L_1}, \baw_{j,L_1})$ have length $\ceil{\log n} +d$ and length at most $(d-1) (\ceil{\log \ceil{\log n}} +1)$, respectively. It follows that in the  loop we replace substrings of length $L_1$ with substrings of length at most 
\begin{align*}
& 4d+d\ceil{\log \ceil{\log n}}+(\ceil{\log n} +d) +(d-1) \ceil{\log(L_1+1)}+\ceil{\log (d+1) }\\
&\leq 4d+d\ceil{\log \ceil{\log n}}+(\ceil{\log n} +d) +(d-1) (\ceil{\log \ceil{\log n}}+1)+\ceil{\log (d+1)}\\
&=L_1-1,
\end{align*}
where the first inequality is obtained by noting that for all sufficiently large $n$ we have $L_1+1\leq 2\ceil{\log n}$. Hence, the loop will execute at most  $\abs{\vw}-L_1+1$ times and the algorithm will terminate eventually.

\begin{algorithm} [ht]
\begin{algorithmic}
\caption{Primal  Pair Elimination Encoder $\cE_2$ for Generating $(L_1,d)$-SD Sequences}
\label{alg:Elim}
\State \textbf{Input}: a $(d\ceil{\log\ceil{\log n}},d)$-WWL sequence $\vw \in \Sigma^{n-K_1-K_2}$
\State \textbf{Output}: a sequence $\baw \in \Sigma^{\leq n-K_1-K_2}$
\State{}
\State Set $\baw = \vw$ and $j_{p} = 0$
\While{ there are two length-$L_1$ substrings in $\baw$ whose Hamming distance is at most $d-1$}
  \State Suppose $(i,j)$ is a primal $(L_1,d-1)$-close window pair in $\baw$ (then necessarily $j>j_p-L_1$)
  \If{$j>  j_{p}-L_1+d$ }
    \State{Remove the substring of length $L_1$ starting at position $j$ and replace it with the sequence
  \[1^d \circ 0^{d\ceil{\log \ceil{\log n}}} \circ 1^d \circ B(i) \circ 1^d \circ \EncDist_{L_1,d-1}(\baw_{i+[L_1]}, \baw_{j+[L_1]})  \circ 0^{\ceil{\log (d+1)}} \circ 1^d  \] }
    \Else
     \State{Set $\baw[j_p+d]$ to be `1'}
     \State{Remove the substring of length $L_1$ starting at position $j$ and replace it  with the sequence
    \[1^d \circ 0^{d\ceil{\log \ceil{\log n}}} \circ 1^d \circ B(i) \circ 1^d \circ \EncDist_{L_1,d-1}(\baw_{i+[L_1]}, \baw_{j+[L_1]})  \circ b(j-j_p+L_1) \circ 1^d \]
    }
  \EndIf
  \State $j_p \gets j$
\EndWhile
\State \Return {$\baw$}
\end{algorithmic}
\end{algorithm}


\begin{lemma} \label{lm:outputofElim}
The output sequence $\baw$ is $(K_1,d)$-WWL and $(L_1,d)$-SD, and the input sequence $\vw$ can be recovered from $\baw$, for all sufficiently large $n$.
\end{lemma}

\begin{IEEEproof}The while loop ensures that the output $\baw$ of  Algorithm~\ref{alg:Elim} is an $(L_1,d)$-SD sequence. Moreover, since $\vw$ is $(d\ceil{\log \ceil{\log n}},d)$-WWL and $K_1= d\ceil{\log \ceil{\log n}} +d$, one can tediously verify that for all large enough $n$, $\baw$ is  $(K_1,d)$-WWL. In particular, even if $\EncDist_{L_1,d-1}(\baw_{i+[L_1]}, \baw_{j+[L_1]})$ is all zeros, for all large enough $n$
\[ K_1-\abs*{\EncDist_{L_1,d-1}(\baw_{i+[L_1]}, \baw_{j+[L_1]})  \circ 0^{\ceil{\log (d+1)}}}\geq d,\]
and a substring of length $K_1$ containing $\EncDist_{L_1,d-1}(\baw_{i+[L_1]}, \baw_{j+[L_1]})\circ 0^{\ceil{\log (d+1)}}$ must also contain at least $d$ of the surrounding $1$'s.

Next, we show  after each replacement the latest inserted substring always starts with the rightmost  $1^d \circ 0^{d\ceil{\log \ceil{\log n}}}$. Let $\baw^{(k)}$ be the sequence $\baw$ after the $k$-th replacement. We prove this by induction. When $k=1$, since $\vw=\baw^{(0)}$ is $(d\ceil{\log \ceil{\log n}},d)$-WWL, the marker $1^d \circ 0^{d\ceil{\log \ceil{\log n}}}$ appears exactly once in $\baw^{(1)}$, and so the claim holds.  Now, in the $k$-th replacement,  $j$ denotes the position of the  substring removed in this replacement, while  $j_p$ denotes the position of the  substring removed  in  the $(k-1)$-th replacement. According to the inductive assumption, the rightmost  $1^d \circ 0^{d\ceil{\log \ceil{\log n}}}$ in $\baw^{(k-1)}$ starts at the position $j_p$.   If $j\geq j_p$, then  the rightmost $1^d \circ 0^{d\ceil{\log \ceil{\log n}}}$ in  $\baw^{(k)}$ is $\baw^{(k)}_{j+[d\ceil{\log \ceil{\log n}}+d]}$. If  $j_{p}-L_1+d <j <j_p$, the overlap of $\baw^{(k-1)}_{j+[L_1]}$ and $\baw^{(k-1)}_{j_p+[L_1]}$ has length greater than $d$. Since the sequence which is inserted in the $k$-th replacement ends with a symbol `1', it can destroy  the marker in $\baw^{(k-1)}_{j_p+[L_1]}$. If   $j_p-L_1 < j \leq  j_{p}-L_1+d$, we set $\baw^{(k)}[j_p+d]$ to be `1' to destroy the  marker in $\baw^{(k-1)}_{j_p+[L_1]}$. In all  cases,  the rightmost   $1^d \circ 0^{d\ceil{\log \ceil{\log n}}}$ in $\baw^{(k)}$ is always  $\baw^{(k)}_{j+[d\ceil{\log \ceil{\log n}}+d]}$.

Now, given the sequence $\baw^{(k)}$, we first search for the rightmost $1^d \circ 0^{d\ceil{\log \ceil{\log n}}}$ in $\baw^{(k)}$ to determine the position $j$. Then from the  substring $\baw^{(k)}_{j+[L_1-1]}$ we can decode $i$, the difference between $\baw^{(k-1)}_{i+[L_1]}$ and $\baw^{(k-1)}_{j+[L_1]}$, and $b(j-j_p+L_1)$. Note that $\baw^{(k-1)}_{i+[\min\set{L_1,j-i}]}= \baw^{(k)}_{i+[\min\set{L_1,j-i}]}$. So we can recover  $\baw^{(k-1)}_{j+[L_1]}$. We remove $\baw^{(k)}_{j+[L_1-1]}$ from $\baw^{(k)}$ and  replace it  with $\baw^{(k-1)}_{j+[L_1]}$. If $ b(j-j_p+L_1)\neq 0^{\ceil{\log (d+1)}}$, we further set the symbol in the position $j_p+d$ to be `0'. In this way, we recover the sequence $\baw^{(k-1)}$. We repeat this procedure until  there is no substring $0^{d\log \log n}$. Then the resulting sequence is the required $\vw$.
\end{IEEEproof}

Now, we need to extend the sequence $\baw$ to a long sequence of length $n$ while keeping the  property of being $(L,d)$-SD.

\begin{lemma}\label{lm:concatenation}
Assume $n$ is sufficiently large. Let $\baw$ be an output of Algorithm~\ref{alg:Elim}. Recall that $K_2=3\ceil{\frac{3}{2}\log L_2}$. 
By invoking Theorem~\ref{thm:singleSDS}  with parameters ``$K=K_2$" and ``$L=L_2$", we get 
  a collection of $(K_2,d)$-WWL sequences $\vs_0,\vs_1,\ldots,\vs_{M-1}$ of length $L_2-L_p$, where $L_p=K_2+d\ceil{\log d}+2d$. Let
\[\bas \eqdef 0^{K_{\max}} \circ \vu \circ \vs_0 \circ 0^{K_{\max}} \circ \vu \circ\vs_1 \circ \cdots \circ 0^{K_{\max}} \circ \vu \circ \vs_{M-1}, \]
where $K_{\max}=\max\set{K_1,K_2}$. Set
\[ \haw  = \cE_3(\baw) \eqdef (\baw \circ 0^{K_2} \circ \bas)[0,n-1].\]
Then $\haw$ is a $(K,d)$-WWL and $(L,d)$-SD sequence  where $K=2(K_1+K_2)$ and $L= \max\{L_1+K_2+K_{\max}+\ell, L_2+2K_1+K_{\max}+\ell\}$. Moreover, $\baw$ can be recovered  from   $\haw$.
\end{lemma}

\begin{IEEEproof}
We first prove that $\bas$ is a $(K_{\max}+K_2,d)$-WWL and $(L_2+K_{\max}-K_2,d)$-SD sequence of length at least $n$. According to the construction, the length of $\bas$  is $M(L_2+K_{\max}-K_2) \geq ML_2$. Recall that $\log M =L_2-3d\log L_2 -7.5 \log L_2 -O(1)$ and $L_2 = \ceil{\log n} +(3d+7)\ceil{\log \ceil{\log n}}$. Then
\begin{align}\label{eq:slen}
ML_2 = 2^{L_2-3d\log L_2 -6.5 \log L_2 -O(1)} =\frac{2^{L_2}}{2^{O(1)} L_2^{3d+6.5} } \geq \frac{n(\log n)^{3d+7}}{ 2^{O(1)}(\log n +(3d+6.5)\log \log n )^{3d+6.5} }>n.
\end{align}
Hence, $\bas$ has length at least $n$. Note that each $\vs_i$ is a $(K_2,d)$-WWL sequence and the length-$d$ prefix of $\vu$ is $1^d$. It follows that $\bas$ is a $(K_{\max}+K_2,d)$-WWL sequence. Moreover, 
note that the sequences $\vs_0,\vs_1,\ldots,\vs_{M-1}$ satisfy the conditions (P1)-(P3) with ``$K=K_2$". 
If $K_2\geq K_1$ (namely, $K_{\max}=K_2$), then by Proposition~\ref{prop:conSDS}, the sequence $\bas$
is an $(L_2,d)$-SD sequence, hence also
an $(L_2+K_{\max}-K_2,d)$-SD sequence. 
If $K_2< K_1$, since the property of being $(K_2,d)$-WWL implies the property of being $(K_{\max},d)$-WWL, the sequences $\vs_0,\vs_1,\ldots,\vs_{M-1}$  also satisfy the conditions (P1)-(P3) with ``$K=K_{\max}$" \footnote{In this case, we take ``$L=L_2+K_{\max}-K_2$", ``$K=K_{\max}$", ``$L_p=K+\ell$", and so, ``$L-L_p=L_2-K_2-\ell$", which is equal to the length of the $\vs_i$'s.}. 
 Again, by Proposition~\ref{prop:conSDS}, the sequence $\bas$ is an $(L_2+K_{\max}-K_2,d)$-SD sequence.

We have shown that  $\bas$ is a $({K_{\max}}+K_2,d)$-WWL sequence in the above paragraph and $\baw$ is a $(K_1,d)$-WWL sequence in Lemma~\ref{lm:outputofElim}. By using the fact that $K_1>d$ and that the $\vu$ substring of $\bas$ starts with $1^d$, it follows that the sequence $\haw=(\baw \circ 0^{K_2} \circ \bas)[0,n-1]$ is $(2(K_1+K_2),d)$-WWL. Now, we shall show that   it  is also $(L,d)$-SD. For any two substrings $\haw_{i+[L]}$ and $\haw_{j+[L]}$ with $i,j \in [n-L+1]$ and $i<j$, we consider the following cases:

\noindent\textbf{Case 1}: $i<j\leq \abs{\baw}-L_1$.   Then
\begin{align*}
d_H(\haw_{i+[L]},\haw_{j+[L]}) \geq d_H(\baw_{i+[L_1]},\baw_{j+[L_1]}) \geq d,
\end{align*}
where the first inequality holds since $L\geq L_1$ and the second inequality holds since  $\baw$  is an $(L_1,d)$-SD sequence.

\noindent\textbf{Case 2}: $i\leq \abs{\baw}-L_1$ and $\abs{\baw}-L_1+1 \leq j \leq \abs{\baw}$. Since $L-L_1\geq K_2+{K_{\max}}+\ell$, where $\ell$ is the length of $\vu$, then $\haw_{j+[L]}$ must contain $0^{K_2+{K_{\max}}}\circ\vu$ as a substring. Assume that $\haw_{j+\delta+[K_2+K_{\max}+\ell]} = 0^{K_2+{K_{\max}}}\circ\vu$ for some $\delta \in [L_1]$.
If $j-i\leq d$, then
\begin{align*}
d_H(\haw_{i+[L]},\haw_{j+[L]}) \geq d_H(\haw_{i+\delta+K_2+{K_{\max}}+[\ell]},\haw_{j+\delta+K_2+{K_{\max}}+[\ell]}) = d_H( 0^{j-i}\circ\vu[0,\ell-(j-i)-1] , \vu) \geq d,
\end{align*}
where the last inequality follows from the definition of a $d$-auto-cyclic sequence.
If $d<j-i \leq K_2+{K_{\max}}$, {since the prefix of $\vu$ is $1^d$,} then
\begin{align*}
d_H(\haw_{i+[L]},\haw_{j+[L]}) \geq d_H(\haw_{i+\delta+K_2+{K_{\max}}+[d]},\haw_{j+\delta+K_2+{K_{\max}}+[d]}) = d_H( 0^d , 1^d) = d.
\end{align*}
If $j-i > K_2+{K_{\max}}$, then $i+\delta+K_2+{K_{\max}}<j+\delta$, and so, $\haw_{i+\delta+[K_2+{K_{\max}}]}$ is a substring of $\baw$. Hence,
\begin{align*}
d_H(\haw_{i+[L]},\haw_{j+[L]}) \geq d_H(\haw_{i+\delta+[K_2+{K_{\max}}]},\haw_{j+\delta+[K_2+{K_{\max}}]}) = d_H( \haw_{i+\delta+[K_2+{K_{\max}}]} ,0^{K_2+{K_{\max}}}) \geq d,
\end{align*}
where the last inequality holds since $\baw$  is a $(K_1,d)$-WWL sequence.

Case 3 and Case 4, which now follow, together cover the case of $i\leq \abs{\baw}-L_1$ and $j> \abs{\baw}$ and the case of $\abs{\baw}-L_1< i < \abs{\baw}$ and $i<j$,

\noindent\textbf{Case 3}: $i\leq \abs{\baw}-(L_2+2K_1-K_2)$ $(\leq \abs{\baw}-L_1)$ and $j >\abs{\baw}$. Denote $L'\eqdef (L_2-K_2)+2K_1$. Then $L\geq L'$. Note that $\haw_{j+[L']}$ always contains $0^{K_1}$ as a substring, and $\haw_{i+[L']}$ is a substring of $\baw$, which is $(K_1,d)$-WWL. Hence,
\begin{align*}
d_H(\haw_{i+[L]},\haw_{j+[L]}) \geq d_H(\haw_{i+[L']},\haw_{j+[L']})  \geq d.
\end{align*}

\noindent\textbf{Case 4}: $\abs{\baw}-(L_2+2K_1-K_2)+1\leq i<\abs{\baw}$ and $i<j$. Since $L\geq (L_2+2K_1-K_2)+K_2+{K_{\max}}+\ell$, $\haw_{i+[L]}$ must contain $0^{K_2+{K_{\max}}} \circ \vu$ as a substring. If $j-i \leq K_2+{K_{\max}}$, then $\haw_{j+[L]}$ must contain $\vu$ as a substring, and so, with the same argument as that in Case 2, one can show that $d_H(\haw_{i+[L]},\haw_{j+[L]}) \geq d$. If $j-i > K_2+{K_{\max}}$, assume that $\haw_{i+\delta'+[K_2+{K_{\max}}]}$ is the all-zero substring  of length $K_2+{K_{\max}}$. Then $j+\delta'>i+\delta'+K_2+{K_{\max}}$. It follows that $\haw_{j+\delta'+[K_2+{K_{\max}}]}$ is a substring of $\bas$, which is $(K_2+{K_{\max}},d)$-WWL. Hence,
\begin{align*}
d_H(\haw_{i+[L]},\haw_{j+[L]}) \geq d_H(\haw_{i+\delta'+[K_2+{K_{\max}}]},\haw_{j+\delta' +[ K_2+{K_{\max}}]}) \geq d.
\end{align*}

\noindent\textbf{Case 5}: $\abs{\baw} \leq i<j$. Then
\begin{align*}
d_H(\haw_{i+[L]},\haw_{j+[L]}) \geq d_H(\haw_{i+K_2+[L-K_2]},\haw_{j+K_2+[L-K_2]}) = d_H(\bas_{i-\abs{\baw}+[L-K_2]},\bas_{j-\abs{\baw}+[L-K_2]}) \geq d,
\end{align*}
where the second inequality holds since $L-K_2\geq L_2 +{K_{\max}}-K_2$ and $\bas$  is $(L_2+{K_{\max}}-K_2,d)$-SD.

Finally, note that in the sequence $\haw$ there is exactly one run of `0' which has length at least  $K_2+K_{\max}$. So we can search for the rightmost $0^{K_2+K_{\max}}$ in $\haw$ and remove this substring as well as the suffix after it to recover the sequence $\baw$.
\end{IEEEproof}

\begin{theorem}\label{thm:SDsequences}
Let $\cE_{\mathtt{SD}}(\cdot) \eqdef  \cE_3(\cE_2(\cE_1(\cdot)))$. Then, for $n$ large enough, $\cE_{\mathtt{SD}}:\Sigma^{n'} \rightarrow \Sigma^n$  is invertible and can encode sequences of $\Sigma^{n'}$ into $(K,d)$-WWL and $(L,d)$-SD sequences where $K=(2d+9)\log \log n +O(1)$  and  
\[
L \begin{cases}
   =\ceil{\log n} +(6d+7) \ceil{\log \ceil{\log n}}+ d \lceil \log d \rceil+5d & \text{if $d\geq 5$}, \\
   \leq \ceil{\log n} +(5d+11.5) \ceil{\log \ceil{\log n}}+ d \lceil \log d \rceil+4d+7.5  & \text{otherwise.}
  \end{cases}
\]
Moreover, $n-n' = \Theta(n/\log n)$, and so,  we have that
\[\lim_{n \to \infty} \frac{n'}{n}  =1.\]
\end{theorem}

\begin{IEEEproof} The statement about $\cE_{\mathtt{SD}}$ follows from Lemma~\ref{lm:outputofElim} and Lemma~\ref{lm:concatenation}. Recall that the encoder $\cE_1$ requires $\Theta(n/\log n)$ redundancies (see \eqref{eq:redundancy}) and $K_1+K_2=\Theta(\log \log n)$. Hence,
\[ n-n'= K_1+K_2+\Theta(n/\log n)= \Theta(n/\log n).  \]
\end{IEEEproof}

\section{Generalized Reconstruction from Noisy Substring Trace}\label{Sec:tracecode}

In this section, we are going to give constructions of $(\Lm,\Lo,e)$-trace reconstruction codes. Our first result generalizes  Proposition~\ref{prop:SDtorecon} and Proposition~\ref{prop:UStorecon}, which shows that the property of being  $(\Lo,d)$-substring distant implies the property of being $(\Lm,\Lo,e)$-trace reconstructible.

\begin{proposition}\label{prop:SDtogenrecon}
Suppose that $\Lm>\Lo$. If a sequence $\vx \in \Sigma^n$ is $(\Lo,4e+1)$-substring distant, then $\vx$ is $(\Lm,\Lo,e)$-trace reconstructible.
\end{proposition}

\begin{IEEEproof}
Let $\cY=\set{\vy^{(0)},\vy^{(1)},\ldots,\vy^{(m-1)}}$ be an $(\Lm,\Lo,e)$-erroneous trace of $\vx$ where the location of each $\vy^{(j)}$ in $\vx$ is $i_j$.
 Since $\vx$ is $(\Lo,4e+1)$-substring distant,  for any two substrings $\vy^{(j)}$ and $\vy^{(j')}$  and their any two subsubstrings $\vy^{(j)}_{k+[\Lo]}$ and $\vy^{(j')}_{k'+[\Lo]}$, we have that
 \begin{equation*}
d_H\parenv*{\vy^{(j)}_{k+[\Lo]},\vy^{(j')}_{k'+[\Lo]}}  \begin{cases} \geq 2e+1   & \text{ if $i_j+k \neq i_{j'}+k'$}, \\
  \leq 2e & \text{ if $i_j+k = i_{j'}+k'$}. \\
\end{cases}
\end{equation*}

Therefore, $\vy^{(0)}$ can be identified as the unique substring $\vy \in \cY$ whose length-$\Lo$ prefix is of Hamming distance at least $2e+1$ from every length-$\Lo$ subsubstring of any other $\vy'\in \cY\backslash\set{\vy}$. Denote the length-$\Lo$ suffix of $\vy^{(0)}$ as $\vs_0$. Then we can identify the substrings $\vy$'s in $\cY$ which overlap  $\vy^{(0)}$ at at least $\Lo$ positions, since each of them contains a unique length-$\Lo$ subsubstring $\vw$ whose distance from $\vs_0$  is at most $2e$. Furthermore, the locations of these substrings in $\vx$ can be determined by aligning the  subsubstring $\vw$ and the suffix $\vs_0$. Assume that there are $m'$ such substrings. Then we have identified the substrings $\vy^{(1)},\ldots,\vy^{(m')}\in \cY$. Next, we consider the length-$\Lo$ suffix of $\vy^{(m')}$ and we can identify all the subsrings in $\cY$ which  overlap  $\vy^{(m')}$ at at least $\Lo$ positions. We repeat the procedure above. Finally, we can determine the location of every substring $\vy\in \cY$ in $\vx$.
\end{IEEEproof}

Combining Theorem~\ref{thm:SDsequences} and Proposition~\ref{prop:SDtogenrecon}, we have the following result.

\begin{corollary}\label{cor:genrecon-1}
Suppose  that $\Lo=\ceil{\log n}+(24e+13)\ceil{\log \ceil{\log n}}+(4e+1)\ceil{\log (4e+1)} +20e+5$ and $\Lm > \Lo$. If $n$ is sufficiently large, then there is an $(\Lm,\Lo,e)$-trace reconstruction code of $\Sigma^n$ whose rate is $1-o(1)$.
\end{corollary}

Now, we consider another parameter regime. Suppose that
\begin{align*}
\Lm&=\ceil{a\log n},\\
\Lo&=\ceil{\gamma \Lm},
\end{align*}
where $a>1$ and $0< a\gamma \leq 1$ are real constants. We are going to construct an $(\Lm,\Lo,e)$-trace reconstruction  code whose rate approaches $\frac{1-1/a}{1-\gamma}$. The basic idea of our code construction is similar to the one in \cite{MarYaa:2021} for the noiseless scenario: A message $\vm$ is encoded into a codeword $\vw=\vw_0\circ \vw_1 \circ \cdots \circ \vw_{2^I-1}$ such that
   \begin{enumerate}
     \item[(i)]  the index $i$ can be decoded from any length-$\Lm$ substring of $\vw_i$ even if the substring is corrupted by at most $e$ errors;
     \item[(ii)] $\vw_i$ can be reconstructed from its any  $(\Lm,\Lo,e)$-erroneous trace.
   \end{enumerate}
To this end, our construction leverages the map $\cE_{\mathtt{SD}}$ in Section~\ref{sec:SDseq} which can encode WWL and SD sequences, as well as the following coded indices $\vc_i$'s which are generated from a robust positioning sequence.

\begin{construction}[Index Construction]
\label{indexconstruction}
Given $e$, let
\begin{align*}
    d_1&\eqdef 2e+1,\\
    d_2&\eqdef 4e+1.
\end{align*}
Additionally, set 
\begin{align*}
I&\eqdef \ceil*{\frac{1-\gamma a}{1-\gamma} \log n+(\log n)^{0.5+\epsilon}},\\
r_I&\eqdef \ceil*{(3d_1+8) \log I},
\end{align*}
where $0<\epsilon<0.5$ is an arbitrary fixed number which is independent of $n$. Then 
\[(I+r_I)-(3d_1+7.5)\log (I+r_I)  - O(1) = I+0.5 \log I -O(1) >I,\]
where we assume $e,a,\gamma,\epsilon$ are constants, and $n\to\infty$. Applying Theorem~\ref{thm:singleSDS} with $L=I+r_I$, there is an explicit construction of sequences  $\vc_0,\vc_1,\ldots,\vc_{2^I-1} \in \Sigma^{I+r_I}$ such that the concatenation
  \[\vc \eqdef \vc_0 \circ \vc_1 \circ \cdots \circ \vc_{2^I-1}\]
   is an $(I+r_I,d_1)$-SD sequence. Moreover, according to the remark following Theorem~\ref{thm:singleSDS}, each $\vc_i$ is $(3\ceil*{\frac{3}{2} \log (I+r_I)}+\ell_{d_1},d_1)$-WWL where
   \[\ell_{d_1}\eqdef d_1 \ceil{\log d_1}+2d_1\]
   is the length of the $d_1$-auto-cyclic sequence $\vu$.   Denote
   \begin{align*}
   K&\eqdef\ceil*{\sqrt{\log n}},\\
   F&\eqdef \ceil*{\frac{I+r_I}{K}}.
   \end{align*}
   For each $i \in [2^I]$, we partition the sequence $\vc_i$ into segments $\vc_i^{(0)},\vc_i^{(1)},\ldots,\vc_i^{(F-1)}$, each of length $\ceil{\frac{I+r_I}{F}}$ or $\floor{\frac{I+r_I}{F}}$. 
\end{construction}

In the following, we first  consider the case of $\Lm\mid n$ and give the code construction. Then we will show how to modify this construction to settle the other cases.

\subsection{The case of $\Lm\mid n$}

Let us define
\begin{align*}
r & \eqdef I+r_I+K+\ell_{d_1}+d_1,\\
L & \eqdef \ceil*{ \parenv*{\Lo-K-\ell_{d_1}-d_1-2\ceil*{\frac{I+r_I}{F}}} \frac{\Lm-r}{ \Lm-r+I+r_I}}.
\end{align*}
We note that by our choice of parameters, $\Lm>r$ for all sufficiently large $n$.
Assume that $\Lm\mid n$ and denote $n_L\eqdef \frac{n}{\Lm}$. 
For each $i \in [2^I]$, let 
 \begin{equation}\label{eq:defNi}
N_i \eqdef \begin{cases} 
\ceil{n_L/2^I} (\Lm-r) & \text{if $i < n_L\bmod 2^I$,} \\
\floor{n_L/2^I} (\Lm-r) & \text{otherwise.} \\
\end{cases}
\end{equation}
Then $\sum_{i\in [2^I]}N_i=n_L(\Lm-r).$

\begin{lemma}\label{lm:precon} Let $K,L,N_i$ be defined as above, and assume $n$ is large enough.
Then for each $i\in [2^I]$ there is an integer $m(N_i)$ with $N_i-m(N_i) = \Theta(N_i/\log N_i)$ and an invertible map  $\cE_{\mathtt{SD}}^{(i)}:\Sigma^{m(N_i)} \rightarrow \Sigma^{N_i}$ which can encode sequences of $\Sigma^{m(N_i)}$ into $(\floor{K/4},d_2)$-WWL and $(L,d_2)$-SD sequences.
\end{lemma}

\begin{IEEEproof}
We shall apply Theorem~\ref{thm:SDsequences} to prove this lemma. To this end, we first need to verify that $N_i$ can be arbitrarily large. As noted before, $\Lm-r>0$. Additionally, $n_L=\Theta(n/\log n)$, and $2^I=n^{\frac{1-\gamma a}{1-\gamma}(1+o(1))}$ and by our choice of parameters, $\frac{1-\gamma a}{1-\gamma}<1$ is a constant. Hence, $N_i\to\infty$ as $n\to\infty$.

Next, we need to verify that $\floor{K/4}$ and $L$ satisfy the two conditions in Theorem~\ref{thm:SDsequences}. Regarding the value of $K$, we need to show that $\floor{K/4}\geq (2d_2+9)\log\log N_i  + O(1)$. Noting that $r_I= \ceil{(3d_1+8)\log I} = O(\log \log n)$ and $K=O(\sqrt{\log n})$, we have that
\begin{align*}
1-\frac{r}{\Lm}&=1-\frac{I+r_I+K+\ell_{d_1}+d_1 }{\Lm} =1-\frac{(\frac{1-\gamma a}{1-\gamma}) \log n + (\log n)^{0.5+\epsilon} +O(\sqrt{\log n})  }{ {a  \log n}+O(1)} \\
& = 1 - \parenv*{\frac{1/a-\gamma}{1-\gamma} + \frac{1}{a (\log n)^{0.5-\epsilon} } + O\parenv*{\frac{1}{\sqrt{\log n}}}}\frac{a\log n}{ a\log n+O(1)}\\
& = 1 - \parenv*{\frac{1/a-\gamma}{1-\gamma} + \frac{1}{a (\log n)^{0.5-\epsilon} } +O\parenv*{\frac{1}{\sqrt{\log n}}}}\parenv*{1-O\parenv*{\frac{1}{\log n}}}\\
& = \frac{1-1/a}{1-\gamma} - \frac{1}{a (\log n)^{0.5-\epsilon} }- O\parenv*{\frac{1}{\sqrt{\log n}}}.
\end{align*}
It follows that
\begin{align*}
\log N_i & = \log \parenv*{ \frac{n_L}{2^I} (\Lm-r)} \pm O(1)  =  \log\parenv*{\frac{n}{2^I}\parenv*{1-\frac{r}{\Lm}}}\pm O(1) \\
&=  \log n - I \pm O(1) =\frac{\gamma a -\gamma}{1-\gamma} \log n - (\log n)^{0.5+\epsilon} \pm O(1). 
\end{align*}
Since
$K=\ceil*{\sqrt{\log n}}$, we have that $\floor{K/4}$ is substantially larger than $(2d_2+9)\log \log N_i +O(1).$

Now, we verify the  condition on $L$, namely that $L\geq \log N_i + (6d_2+7)\log \log N_i +O(1)$. Note that
\begin{align*}
\frac{I+r_I}{\Lm-r}&  =\frac{I+O(\log {\log n})}{\Lm-I -O(\sqrt{\log n})} = \frac{I}{\Lm-I} \cdot \frac{1+O(\log \log n/{\log n})}{1-O(1/\sqrt{\log n}) } \\
&= \frac{I}{\Lm-I}\parenv*{1+O\parenv*{\frac{1}{\sqrt{\log n}} }   }  = \frac{I}{\Lm-I}+O\parenv*{\frac{1}{\sqrt{\log n}}},
\end{align*}
and
\begin{align*}
\ceil*{ \frac{I+r_I}{F}}=\ceil*{ \frac{I+r_I}{\ceil{(I+r_I)/K}}}\leq \frac{I+r_I}{(I+r_I)/K}+1=K+1.
\end{align*}

Hence, we have that
\begin{align*}
L & =  \ceil*{ \parenv*{\Lo-K-\ell_{d_1}-d_1-2\ceil*{\frac{I+r_I}{F}}} \frac{\Lm-r}{ \Lm-r+I+r_I}} \\ 
& \geq \frac{\Lo-3K-\ell_{d_1}-d_1-2}{1+(I+r_I)/(\Lm-r)}=\frac{\Lo - O(\sqrt{\log n})}{ \frac{\Lm}{\Lm-I}+O\parenv*{{1}/{\sqrt{\log n}}} }\\
& =\frac{\Lo(\Lm-I)}{\Lm} \cdot \frac{1-O(1/\sqrt{\log n})}{1+O(1/\sqrt{\log n}) }\\
& \geq \gamma \parenv*{a\log n -\frac{1-\gamma a}{1-\gamma} \log n - (\log n)^{0.5+\epsilon} -1}  \parenv*{1-O\parenv*{\frac{1}{\sqrt{\log n}} }   }   \\
&= \frac{\gamma a -\gamma }{1-\gamma } \log n -\gamma (\log n)^{0.5+\epsilon}-O(\sqrt{\log n}).
\end{align*}
It follows that
\begin{align*}
L-\log N_i= (1-\gamma) (\log n)^{0.5+\epsilon} -O(\sqrt{\log n}) = \omega(\log \log N_i).
\end{align*}
We can conclude that $L$ is substantially larger than $\log N_i + (6d_2+7)\log \log N_i +O(1)$.
\end{IEEEproof}

 Now, we present our code construction.

\begin{construction}
\label{Construction1} Let $m(N_i)$'s be defined as in Lemma~\ref{lm:precon}. We now describe a mapping from $\Sigma^{\sum_{i\in [2^I]} m(N_i)}$ to $\Sigma^n$. For any message $\vm \in \Sigma^{\sum_{i\in [2^I]} m(N_i)}$, partition $\vm$ into $2^I$ substrings:
\[\vm = \vm_0\circ \vm_1 \circ \cdots \circ \vm_{2^I-1},\]
where each $\vm_i$ has length $m(N_i)$.
For each  $i \in [2^I]$, let
\[\vv_i = \cE_{\mathtt{SD}}^{(i)}(\vm_i) \in \Sigma^{N_i},\]
where $\cE_{\mathtt{SD}}^{(i)}$ is the map mentioned in Lemma~\ref{lm:precon}.
We partition each $\vv_i$ into substrings of  length $\Lm-r$:
\begin{equation*}
\vv_i =  \begin{cases} 
\vv_{i,0} \circ \vv_{i,1} \circ \cdots \circ \vv_{i,\ceil{n_L/2^I}-1} &\text{if $i < n_L\bmod 2^I$}, \\
 \vv_{i,0} \circ \vv_{i,1} \circ \cdots \circ \vv_{i,\floor{n_L/2^I}-1} &\text{otherwise}. \\
\end{cases}
\end{equation*}
Then the total number of $\vv_{i,j}$'s is  $n_L$.
We further partition each  $\vv_{i,j}$ into $F$ segments of lengths $\ceil{(\Lm-r)/F}$ or $\floor{(\Lm-r)/F}$:
\[ \vv_{i,j} = \vv_{i,j}^{(0)} \circ \vv_{i,j}^{(1)} \circ \cdots \circ \vv_{i,j}^{(F-1)}.\]
Recall $\vc_{i}^{(m)}$ from the index construction, Construction~\ref{indexconstruction}. Let
\begin{equation*}
\vw_{i,j} \eqdef  \begin{cases} 
0^{d_1}  \circ  \vv_{i,j}^{(0)} \circ    \vc_{i}^{(0)}  \circ \cdots \circ  \vv_{i,j}^{(F-1)} \circ    \vc_{i}^{(F-1)}  &    \text{if $j=0$}, \\
1^{d_1}  \circ  \vv_{i,j}^{(0)} \circ    \vc_{i}^{(0)}  \circ \cdots \circ  \vv_{i,j}^{(F-1)} \circ    \vc_{i}^{(F-1)} & \text{otherwise}. \\
\end{cases}
\end{equation*}

Finally, let 
 \begin{equation*}
\vw_i = \begin{cases}
\vp \circ \vw_{i,0} \circ\vp \circ \vw_{i,1} \circ \cdots \circ \vp \circ \vw_{i,\ceil{n_L/2^I}-1} & \text{if $i < n_L\bmod 2^I$}, \\
\vp \circ \vw_{i,0} \circ\vp \circ \vw_{i,1} \circ \cdots \circ \vp \circ \vw_{i,\floor{n_L/2^I}-1} & \text{otherwise}, \\
\end{cases}
\end{equation*}
where $\vp\eqdef 0^K \circ \vu$ and  $\vu$ is the $d_1$-auto-cyclic sequence in Theorem~\ref{thm:acv}. 
Denote
\[\vw \eqdef  \vw_{0}\circ \vw_{1} \circ \cdots  \circ \vw_{2^I-1}. \]
The constructed code, $\cCt$, is the image of the mapping described above.
\end{construction}

\begin{lemma}\label{lm:coderate}
Let $\cCt$ be the code obtained by Construction~\ref{Construction1}. Then $\cCt \subseteq \Sigma^n$ and its rate is
\[ R(\cCt) =  \frac{1-1/a}{1-\gamma} - \frac{1}{a (\log n)^{0.5-\epsilon} } - O\parenv*{\frac{1}{\sqrt{\log n}}}.\]
\end{lemma}
\begin{IEEEproof}
In our  construction, every sequence $\vw_{i,j}$ has length $\Lm-r+d_1+\abs{\vc_i}=\Lm-K-\ell_{d_1}$, and so, the concatenation  $\vp \circ \vw_{i,j}$ has length $\Lm$. It follows that the codeword $\vw$ has length $n_L \Lm  =n$. 
 Noting that the map $\cE_{\mathtt{SD}}$ is invertible, we can uniquely recover $\vm$ from $\vw$. Therefore, the code $\cCt$ has rate $\sum_{i\in[2^I] }m(N_i)/{n}$. 

We have shown in the proof of  Lemma~\ref{lm:precon}  that
\[1-\frac{r}{\Lm}= \frac{1-1/a}{1-\gamma} - \frac{1}{a (\log n)^{0.5-\epsilon} }- O\parenv*{\frac{1}{\sqrt{\log n}}},\]
and for each $i \in [2^I]$,
\[\log N_i =\Theta(\log n).\]
Hence,
\begin{align*}
R(\cCt)&=\frac{\sum_{i \in [2^I]} m(N_i)}{n} =\frac{\sum_{i \in [2^I]} N_i-\Theta\parenv*{N_i/\log N_i} }{n} \\
& =\frac{\sum_{i \in [2^I]} N_i }{n}\parenv*{1-\Theta\parenv*{\frac{1}{\log n}}} = \frac{n_L(\Lm-r) }{n}\parenv*{1-\Theta\parenv*{\frac{1}{\log n}}}\\
& = \parenv*{1-\frac{r}{\Lm}} \parenv*{1-\Theta\parenv*{\frac{1}{\log n}}} =  \frac{1-1/a}{1-\gamma} - \frac{1}{a (\log n)^{0.5-\epsilon} } - O\parenv*{\frac{1}{\sqrt{\log n}}}. 
\end{align*}
\end{IEEEproof}

In the following, we shall show that the code $\cCt$  is an $(\Lm,\Lo,e)$-trace reconstruction  code. 

\begin{lemma}[Construction~1 and Lemma~3.6 in \cite{Cheeetal:2020}]\label{lm:pcompare}
Let $\vw= \vp \circ \vw_{0,0} \circ\vp \circ \vw_{0,1} \circ \cdots \circ \vp \circ \vw_{2^I-1,\floor{n_L/2^I}-1 }  $ be a codeword of $\cCt$. Assume that
the substrings $\vw_{i,j}$'s  satisfy the following conditions:
\begin{enumerate}
\item[(P1)] $\vw_{i,j}$ is a $(K,d_1)$-WWL sequence for each $(i,j)$; and 
\item[(P2)] $\vw_{i,j}[0,\mu-1] \circ \vw_{i',j'}[\mu,\Lm-K-\ell_{d_1}-1]$ is a $(K,d_1)$-WWL sequence for $(i,j),(i',j')$ such that $(i,j)\neq (i',j')$ and $\mu\in [\Lm-K-\ell_{d_1}]$.
\end{enumerate}
Then for every substring $\vy=\vw_{i_0+[\Lm]}$ in $\vw$ and each\footnote{If $i \in [\Lm-K+\ell_{d_1},\Lm-1 ]$, we let $\vy_{i+[K+\ell_{d_1}]}$ denote the concatenation $\vy[i,\Lm-1]\circ \vy[0,K+\ell_{d_1}-(\Lm-i)-1]$. } $i\in [\Lm]$, the following hold:
\begin{enumerate}[(i)]
\item If $i+i_0\equiv 0 \ppmod{\Lm}$, then $\vy_{i+[K+\ell_{d_1}]}=\vp$.
\item If $i+i_0\not \equiv 0 \ppmod{\Lm}$, then $d_H(\vy_{i+[K+\ell_{d_1}]},\vp)\geq d_1$.
\end{enumerate}
\end{lemma}

\begin{lemma}\label{lm:index}
Assume $n$ is sufficiently large. Let $\vy$ be an arbitrary length-$\Lm$ substring of $\vw \in \cCt$. Then $\vy$ contains a length-$(I+r_I-\mu)$ suffix of  a coded index $\vc_i$ and a length-$\mu$ prefix of either $\vc_i$ or $\vc_{i+1}$ for some $i\in [2^{I}]$ and $\mu \in [I+r_I]$. Furthermore, even if  $\vy$ is corrupted by at most $e$ errors,  we can still identify the positions  where the  said suffix and prefix appear, and so reconstruct them with at most $e$ errors.
\end{lemma}

\begin{IEEEproof} 
We note that the length of $\vp\circ\vw_{i,j}$ is $\Lm$, and that $\vw$ is a concatenation of such strings. Hence, the first statement follows directly from the code construction. Now, assume that $\vy$ is corrupted by at most $e$ errors. We shall use Lemma~\ref{lm:pcompare} to identify the location of the marker $\vp$ in $\vy$.
Recall that every $\vc_i$ is $(3\ceil*{\frac{3}{2} \log (I+r_I)}+\ell_{d_1},d_1)$-WWL (see the index construction, Construction~\ref{indexconstruction}) and every $\vv_i$ is $(\floor{K/4},d_2)$-WWL (see Lemma~\ref{lm:precon}). Since $3\ceil*{\frac{3}{2} \log (I+r_I)}+\ell_{d_1}<\floor{K/4}$ and $d_1<d_2$, all the segments $\vc_{i}^{(h)}$'s and  $\vv_{i,j}^{(h)}$'s are  $(\floor{K/4},d_1)$-WWL. Hence, $\vw_{i,j}$'s satisfy the conditions in Lemma~\ref{lm:pcompare}. This follows since any substring of length $K$ contains a substring of length $\floor{K/4}$ that is fully contained within a segment of the form $\vc_{i}^{(h)}$ or  $\vv_{i,j}^{(h)}$, thus providing the minimum weight of $d_1$ as claimed.

Since $\vy$ suffers from at most $e$ errors and $d_1=2e+1$, by Lemma~\ref{lm:pcompare} there is a unique index $i \in [\Lm]$ such that
\[d_H(\vy_{i+[K+\ell_{d_1}]}, \vp) \leq e.\]
Hence, by comparing the distance between the marker $\vp$ and each length-$\ell_p$ substring of $\vy$, we can identify the location of  the marker in $\vy$. Once the marker $\vp$ is located, the positions in which the symbols of the coded indices $\vc_{i}^{(h)}$'s appear  can also be determined. Then we can reconstruct a prefix  $\vc_i[\mu,I+r_I-1]$ and a suffix  $\vc_i[0,\mu-1]$ or $\vc_{i+1}[\mu-1]$ for some $\mu \in [I+r_I]$ with at most $e$ errors.
\end{IEEEproof}

The following lemma ensures that every length-$\Lo$ substring of $\vw$ contains a long-enough substring of the $(L,d_2)$-SD sequence $\vv_i$. 

\begin{lemma}\label{lm:overlap}
Assume $n$ is sufficiently large. Let $\vw$ be a codeword of $\cCt$. Then every length-$\Lo$ substring of $\vw$ contains at least $L$ consecutive symbols of $\vv =\vv_0\circ \vv_1\circ \cdots \circ \vv_{2^I-1}$.
\end{lemma}

\begin{IEEEproof}
Note that the concatenation
\[ \vv_{i,j}^{(0)} \circ    \vc_{i}^{(0)}  \circ \cdots \circ  \vv_{i,j}^{(F-1)} \circ    \vc_{i}^{(F-1)}\]
consists of $\abs{\vv_{i,j}}+\abs{\vc_i}=\Lm-r+I+r_I$ symbols, out of which $\abs{\vv_{i,j}}=\Lm-r$ symbols are from $\vv$. Then according to the construction,  every length-$\Lo$ substring of $\vw$ contains at least 
\[\parenv*{\Lo-(K+\ell_{d_1})-d_1-2\ceil*{\frac{I+r_I}{F}} 
 }\frac{\Lm-r}{\Lm-r+I+r_I}\]
consecutive symbols of $\vv$, where $\Lo-(K+\ell_{d_1})-d_1-2\ceil*{\frac{I+r_I}{F}}$ accounts for the worst case where the substring  both begins  and ends  with some segments of the coded indices (of length $\ceil*{\frac{I+r_I}{F}}$ or $\floor*{\frac{I+r_I}{F}}$ ) and contains a copy of $\vp\circ 0^{d_1}$ or $\vp\circ 1^{d_1}$.  
\end{IEEEproof}

\begin{theorem}\label{thm:tracerecon}
The code $\cCt$ obtained in Construction~\ref{Construction1} is an $(\Lm,\Lo,e)$-trace reconstruction code of $\Sigma^n$ with  rate
\[ R(\cCt) =  \frac{1-1/a}{1-\gamma} - \frac{1}{a (\log n)^{0.5-\epsilon} } - O\parenv*{\frac{1}{\sqrt{\log n}}}.\]
\end{theorem}

\begin{IEEEproof} The code rate  has been calculated in Lemma~\ref{lm:coderate}.
Let $\vw$ be a codeword of $\cCt$ and $\cY$ be an $(\Lm,\Lo,e)$-erroneous trace of $\vw$. For each $\vy$ in $\cY$, since the length of $\vy$ is at least $\Lm$, according to Lemma~\ref{lm:index}, we can extract a corrupted copy $\vc_{\suf}$ of the length-$(I+r_I-\mu)$ suffix of $\vc_i$, and a corrupted copy $\vc_{\pre}$ of a length-$\mu$ prefix of either $\vc_i$ or $\vc_{i+1}$, with the total number of errors being no more than $e$. Consider the following cases.
\begin{enumerate}
\item If $\mu=0$,  then $\vc_{\suf}$ is a corrupted copy of $\vc_i$, and so, we can run the locating algorithm of the robust positioning sequence $\vc=\vc_0\circ \vc_1\circ \cdots \circ \vc_{2^I-1}$ on the corrupted $\vc_{\suf}$ to determine the index $i$.
\item If $\mu > 0$ then $\vy$ contains a copy of either $\vp\circ 0^{d_1}$ or $\vp \circ 1^{d_1}$ with at most $e$ errors. Since $d_1=2e+1$, we can distinguish these two cases.
    \begin{enumerate}
     \item If $\vy$ contains a copy of $\vp\circ 0^{d_1}$, then $\vc_{\pre}$ is a prefix of $\vc_{i+1}$, and so, we  run the locating algorithm of $\vc$ on $\vc_{\suf} \circ \vc_{\pre}$ to decode the index $i$.
     \item If $\vy$ contains a copy of $\vp\circ 1^{d_1}$, then $\vc_{\pre}$ is a prefix of $\vc_{i}$, and so, we  run the locating algorithm of $\vc$ on $\vc_{\pre} \circ \vc_{\suf}$ to decode the index $i$.
     \end{enumerate}
\end{enumerate}

The discussion above shows that for every string $\vy\in \cY$, we can decode the index $i$. If $\vy$ intersects both $\vv_i$ and $\vv_{i+1}$, then we can determine its location in $\vw$ by identifying the location of the marker $\vp$ in $\vy$. For the other strings with index $i$, since $\vv_i$ is an $(L,4e+1)$-SD sequence, according to Lemma~\ref{lm:overlap} and Proposition~\ref{prop:SDtogenrecon},  there is a unique way to determine the correct order of these strings and match correctly the suffix and the prefix of consecutive strings. By taking the majority value at every position, we can reconstruct a sequence $\vw_i'$, which is a long substring of $\vw_i$ possibly with some errors. It remains to determine the location of $\vw_i'$ in $\vw_i$, which can be done as follows.
\begin{enumerate}
 \item If $\vw_i'$ contains a corrupted copy of $\vp \circ 0^{d_1}$ with at most $e$ errors, then the location this marker in $\vw_i'$ determines the location of $\vw_i'$ in $\vw_i$, since $\vw_i$ only contains one copy of $\vp\circ 0^{d_1}$.
 \item If $\vw_i'$ does not contain any corrupted copy of $\vp \circ 0^{d_1}$ up to $e$ errors, then there is a string $\hay\in \cY$ which intersects both $\vw_{i-1}$ and $\vw_i$ and contains  $\vp \circ 0^{d_1}$ as a substring  with at most $e$ errors,  since the length of $\vp \circ 0^{d_1}$ is less that $\Lo$. 
   \begin{enumerate}
      \item If $\hay$ overlaps $\vw_i$ in at most $\Lo$ positions, since $\Lo <\Lm$, $\vw_i'$ must contain a copy of the first  $\vp\circ 1^{d_1}$ of $\vw_i$, and so,  the location of $\vw_i'$ in $\vw_i$ can be determined by identifying  the first occurrence  of the marker $\vp$ in $\vw_i'$.
      \item If $\hay$ overlaps $\vw_i$ in at least $\Lo$ positions, then $\hay$ and the length-$\Lo$ prefix of $\vw_i'$ share a length-$L$ substring of $\vv_i$. Since $\vv_i$ is $(L,4e+1)$-SD, we can match the suffix of $\hay$ and the prefix of $\vw_i'$ correctly. Then the location of $\vw_i'$ in $\vw$ can be deduced from the location of $\hay$ in $\vw$.
   \end{enumerate}
 \end{enumerate}
 \end{IEEEproof}

\subsection{The case of $\Lm\nmid n$}

Now, we consider the case that $\Lm$ does not divide $n$. Take $n_L=\floor{n/\Lm}$. Construction~\ref{Construction1} can yield a trace reconstruction code of block length $n_L \Lm$. Our approach is to extend this code to have length $n$. 
Let $N_i$ be defined as in \eqref{eq:defNi} and $m(N_i)$ be defined as in Lemma~\ref{lm:precon}. For any message $\vm \in \Sigma^{\sum_{i\in [2^I]} m(N_i)}$, partition $\vm$ into $2^I$ substrings, each of length $m(N_i)$:
\[\vm = \vm_0\circ \vm_1 \circ \cdots \circ \vm_{2^I-1}.\]
For each  $i \in [2^I-1]$, let
\[\vv_i = \cE_{\mathtt{SD}}^{(i)}(\vm_i) \in \Sigma^{N_i}.\]
The main difference from the previous case is the encoding of $\vm_{2^I-1}$.
We recall that the encoder $\cE_{\mathtt{SD}}^{(i)}$ first encodes the message $\vm_i$ to an SD and WWL sequence  of length probably less than $N_i$. Then it extends the sequence by appending a sequence $\bas$ and taking the first $N_i$ bits of the concatenation. For $i=2^I-1$, we modify the encoder $\cE_{\mathtt{SD}}^{(2^I-1)}$ by taking the first $N_{2^I-1}+{\Lm-r}$ bits of the concatenation. This is possible since asymptotically the length of $\bas$ is larger than $N_{2^I-1}+{\Lm-r}$, see \eqref{eq:slen}. We denote this modified encoder as $\cE_{\mathtt{SDE}}^{(2^I-1)}$ and let 
\[\vv_{2^I-1}=\cE_{\mathtt{SDE}}^{(2^I-1)}(\vm_{2^I-1}).\]
 Then $\vv_{2^I-1}$ is $(\floor{K/4},d_2)$-WWL and $(L,d_2)$-SD and has length $N_{2^I-1}+\Lm-r=\ceil{n_L/2^I}(\Lm-r)$. Moreover, the message $\vm_{2^I-1}$ can be decoded from the first $N_{2^I-1}$ bits of $\vv_{2^I-1}$. In other words, the last $\Lm-r$ bits are redundant. 

Then, we proceed similarly as in Construction~\ref{Construction1} and obtain an $(\Lm,\Lo,e)$-trace reconstruction code of block length ${(n_L+1)\Lm}$. Note that the last $\Lm$ bits are redundant, and so, we  delete ${(n_L+1)\Lm}-n$  of them to form an $(\Lm,\Lo,e)$-trace reconstruction code of length $n$, with code rate 
\[ \frac{\sum_{i\in [2^I]} m(N_i)}{n} =\parenv*{\frac{1-1/a}{1-\gamma}-o(1)}\frac{n_L \Lm}{n}=\frac{1-1/a}{1-\gamma}-o(1). \]

\subsection{Handling  noise which occurs before sequencing}

Up to now, we have studied $(\Lm,\Lo,e)$-trace reconstruction codes, which allow reconstructing the  maximum reconstructible-string from an erroneous trace $\cY$ of a codeword $\vw$. We use $M(\cY)$ to denote the  maximum reconstructible-string of $\cY$. If $\cY$ is reliable, then $M(\cY)= \vw$. However, if $\cY$ is not reliable, then $M(\cY)$ might be different from $\vw$. This may happen especially when the sequence $\vw$ is subject to errors before its substrings are sampled.   In the remainder of this section, we shall modify Construction~\ref{Construction1} to combat such errors.

Let $\cY$ be an $(\Lm,\Lo,e)$-erroneous trace of $\vw$ such that $d_H(M(\cY), \vw) \leq \tau$, which is referred to as an \emph{$(\Lm,\Lo,e,\tau)$-erroneous trace}. We aim to reconstruct $\vw$ from $\cY$, and so retrieve the message which is stored in $\vw$. Our construction, which is presented below, borrows the idea from \cite[Construction B]{BarMarYaaYeh:22}.

\begin{construction}
\label{Construction2}
Assume that $\Lm \mid n$ and take $n_L=n/\Lm$. Let $N\eqdef \floor{n_L/2^I}(\Lm-r)$. According to Lemma~\ref{lm:precon}, there is an integer $m(N)$ with $N-m(N) = \Theta(N/\log N)$ and an invertible map  $\cE_{\mathtt{SD}}:\Sigma^{m(N)} \rightarrow \Sigma^{N}$ which can encode sequences of $\Sigma^{m(N)}$ into $(\floor{K/4},d_2)$-WWL and $(L,d_2)$-SD sequences. Let $\cE_{\mathtt{SDE}}:\Sigma^{m(N)} \rightarrow \Sigma^{N+\Lm-r}$ be an encoder which modifies $\cE_{\mathtt{SD}}$ by taking the first $N+\Lm-r$ bits of the concatenation. 

For any message $\vm \in \Sigma^{(2^I-2\tau)m(N)}$, we first use a $[2^I,2^I-2\tau,2\tau+1]_{2^{m(N)}}$ Reed-Solomon code\footnote{The Reed-Solomon code is over the finite field of size $2^{m(N)}$. The message is partitioned into groups of $m(N)$ bits, and each group is translated to a single symbol from the finite field. After encoding the reverse translation to bits is performed. Note that $m(N)=N-\Theta(N/\log N)$, $\log (N)=\Theta(\log n)$ and $I=O(\log n)$. Hence, $m(N)>I$ and so, the Reed-Solomon code exists.} to encode $\vm$ into a codeword $\bam \in \Sigma^{{2^I}m(N)}$. 
We partition $\bam$ into sequences of length $\Lm-r$:
\[\bam=\bam_0\circ \bam_1\circ \cdots \circ \bam_{2^I-1}.\]
For each $i \in [2^I]$, let  
\begin{equation*}
\vv_i \eqdef \begin{cases}
\cE_{\mathtt{SDE}}(\bam_i) \in \Sigma^{N+\Lm-r} & \text{if $i < n_L\bmod 2^I$}, \\
\cE_{\mathtt{SD}}(\bam_i)\in \Sigma^{N} & \text{otherwise}. \\
\end{cases}
\end{equation*}

Then we proceed similarly as in Construction~\ref{Construction1} to obtain a sequence $\vw$ of length $n$. We use $\chCt$ to denote the code produced by this construction.
\end{construction}

\begin{lemma}
Let $\vw$ be a codeword of $\chCt$ and $\cY$ be an $(\Lm,\Lo,e,\tau)$-erroneous trace of $\vw$. Then we can recover $\vm$ from $\cY$.
\end{lemma}

\begin{IEEEproof} With the same argument as the proof of Theorem~\ref{thm:tracerecon}, we can show that $\chCt$ is  an $(\Lm,\Lo,e)$-trace reconstruction code of $\Sigma^n$. Since $\cY$ is also an  $(\Lm,\Lo,e)$-erroneous trace of $\vw$, the maximum reconstructible-substring $M(\cY)$ can be decoded from $\cY$.  By reversing the operations in Construction~\ref{Construction2},  we obtain a sequence $\bam' \in \Sigma^{{2^I}m(N)}$  from $M(\cY)$. We partition  $\bam'$ into $2^I$ segments of the same length, i.e., $\bam'=\bam_0'\circ \bam_1'\circ \cdots \circ \bam_{2^I-1}'$. Since  $d_H(M(\cY), \vw) \leq \tau$, then there are at most $\tau$ indices $i\in [2^I]$ such that $\bam_i\neq \bam_i'$. Hence, we can run the decoder of the Reed-Solomon code on $\bam'$ to recover $\bam$.
\end{IEEEproof}

\begin{theorem}
Suppose that $\tau = O\parenv*{n^{\frac{1-\gamma a}{1-\gamma}}}$.
Then the code $\chCt$ obtained in Construction~\ref{Construction2} is an $(\Lm,\Lo,e,\tau)$-trace reconstruction code of $\Sigma^n$ with  rate
\[ R(\chCt) =  \frac{1-1/a}{1-\gamma} -o(1).\]
\end{theorem}

\begin{IEEEproof} Since $\tau = O\parenv*{n^{\frac{1-\gamma a}{1-\gamma}}}$, we have that $2\tau /2^I = o(1)$. Hence, the code rate
\begin{align*}
R(\chCt) & = \frac{(2^I-2\tau)m(N)}{n} =\frac{2^Im(N)}{n}-\frac{2\tau N}{n}\parenv*{1-\Theta\parenv*{\frac{1}{\log N}}}  \\
 &\geq  \frac{2^Im(N)}{n}- \frac{2\tau}{2^I} \parenv*{1-\frac{r}{\Lm}}\parenv*{1-\Theta\parenv*{\frac{1}{\log N}}}\\
 & =  \frac{2^Im(N)}{n}-o(1) .
\end{align*}
Consider the $N_i$'s which are defined in \eqref{eq:defNi}. 
We have that 
\[
N_i \eqdef \begin{cases}
N+ \Lm-r & \text{if $i < n_L\bmod 2^I$}, \\
  N & \text{otherwise}. \\
\end{cases}
\]
Hence,
\begin{align*}
R(\chCt) &  =  \frac{2^Im(N)}{n}-o(1) \\
 &\geq  \frac{\sum_{i\in [2^I]} m(N_i) -2^I(\Lm-r)}{n}- o(1)\\
 & = R(\cCt)-o(1) = \frac{1-1/a}{1-\gamma} -o(1).
\end{align*}
\end{IEEEproof}

\subsection{$(\Lm,0,e)$-Reconstruction Codes}
In this subsection, we consider the case of $\Lo=0$.

\begin{construction}
\label{Construction1prime}
Suppose that $\Lm=\ceil{a \log n }$, $\Lo=0$ and $\Lm \mid n$. As before, we denote $n_L\eqdef \frac{n}{\Lm}$ and $K\eqdef \ceil*{\sqrt{\log n}}$. However, this time, we let $I\eqdef \ceil{\log n_L}$ and $r_I\eqdef \ceil{(3d+8)\log I}$ where $d=2e+1$ and $\ell=d\ceil{\log d}+2d$. Then according to Theorem~\ref{thm:singleSDS}, there is a collection of $(3\ceil{\frac{3}{2}\log (I+r_I)}+\ell,d)$-WWL sequences $\vc_0,\vc_1,\ldots,\vc_{2^I-1}\in \Sigma^{I+r_I}$  such that the concatenation $\vc_0\circ \vc_1\circ \cdots \circ \vc_{2^I-1}$ is an $(I+r_I,d)$-SD sequence.
 
Denote $m'\eqdef  \Lm-(I+r_I+K+\ell)$. Let $\cE_{\mathtt{WWL}}$ be the encoder in \cite[Algorithm~2]{LevYaa:2017} which can encode sequences of $\Sigma^{m'-d}$ into $(\ceil{K/4},d)$-WWL sequences\footnote{Note that $m'=\Theta(\log n)$ and $K=\ceil*{\sqrt{\log n}}$. Hence, $K/4 \gg \cF(m',d)= \log m' + (d-1)\log \log m'+O(1)$. Then according to Lemma 19 in \cite{LevYaa:2017}, the encoder $\cE_{\mathtt{WWL}}$ does work. }  of $\Sigma^{m'}$. For a message $\vm=\vm_0\circ \vm_1\circ \cdots \circ \vm_{n_L-1}$ where $\vm_i\in \Sigma^{m'-d}$ for $i \in[n_L]$,  let $\vw_i\eqdef \cE_{\mathtt{WWL}}(\vm_i) $ for all $i \in [n_L]$. 

Denote $\vp \eqdef 0^K \circ \vu$ where $\vu$ is a $d$-auto-cyclic sequence of length $\ell$. Let
\[\vw = \vp  \circ \vc_{0} \circ  \vw_{0} \circ \vp  \circ \vc_{1} \circ  \vw_{1} \circ \cdots \circ \vp  \circ \vc_{n_L-1} \circ  \vw_{n_L-1}.  \]
Output $\vw$ as the codeword which encodes the message  $\vm$. The image under this mapping is the code that we construct.
\end{construction}

\begin{theorem}\label{thm:gamma0}
The code obtained in Construction~\ref{Construction1prime} is an $(\Lm,0,e)$-trace reconstruction code of $\Sigma^n$ with rate 
\[1-\frac{1}{a}-O\parenv*{\frac{1}{\log n}}.\]
\end{theorem}

\begin{IEEEproof}[Sketch of proof] 
The code has rate
\[ \frac{n_L(m'-d)}{n}=\frac{m'-d}{\Lm}=\frac{\Lm-(I+r_I+K+\ell+d)}{\Lm}=1-\frac{1}{a}-O\parenv*{\frac{1}{\log n}}.\]

Now, let  $\vy$  be a length-$\Lm$ substring of some codeword $\vw$.  Then $\vy$ must contain either a copy of $\vp\circ \vc_{i}$ or a suffix of $\vp\circ \vc_{i}$ together with a prefix of $\vp\circ \vc_{i+1}$. Since  $\vw_i$'s and $\vc_j$'s are WWL sequences, even if $\vy$ suffers from $e$ errors, we can still  locate the marker $\vp$ in $\vy$. Then we can run the locating algorithm of the robust positioning sequence  $\vc_0\circ \vc_1\circ \cdots \circ \vc_{2^I-1}$  to determine the index $i$ or $i+1$, and hence the location of $\vy$.
\end{IEEEproof}

For the case of $\Lm\nmid n$, let $n_L=\ceil{n/\Lm}$. We first construct an $(\Lm,0)$-trace reconstruction code of $\Sigma^{n_L\Lm}$, where the  length-$\Lm$ suffix of every codeword is fixed. Then we truncate it to be of length $n$. In this way, we get a code of rate
\[\frac{\floor{n/\Lm}(\Lm-(I+r_I+K+\ell+d)) }{n}\geq  \parenv*{1-\frac{\Lm-1}{n}} \parenv*{1-\frac{I+r_I+K+\ell+d}{\Lm}}=1-\frac{1}{a}-O\parenv*{\frac{1}{\log n}}.\]

For $(\Lm,0,e,\tau)$-erroneous trace reconstruction, we proceed similarly as  in \cite[Construction B]{BarMarYaaYeh:22}. We first use an $(n_L, 2^{(m'-d)(n_L-r)},2\tau+1)_{2^{m'-d}}$ code to encode a message $\vm=\vm_0\circ \vm_1\circ \cdots \circ \vm_{n_L-r-1} \in \Sigma^{(m'-d)(n_L-r)}$ to a sequence $
\bam=\bam_0\circ \bam_1\circ \cdots \circ \bam_{n_L-1} \in \Sigma^{(m'-d)n_L}$. Then we use the encoder outlined in Construction~\ref{Construction1prime} to get a codeword $\vw$. We note that Construction B in \cite{BarMarYaaYeh:22} only concerns errors before sequencing, while our construction incorporates errors both before and after sequencing. 

\section{Multi-Strand Reconstruction}
\label{sec:mulstrrecon}

In this section, instead of reconstructing a single sequence, we consider the problem of reconstructing a \emph{multiset} of $k$ sequences of length $n$ from the union of their traces.
The following construction of multi-strand $(\Lm,\Lo,e)$-trace reconstruction codes is adapted from \cite[Construction C]{YehBarMarYaa:2023}.

\begin{construction}
\label{Construction3}
Let $N\eqdef k(n-\Lo)+\Lo$. We take an  $(\Lm,\Lo,e)$-trace reconstruction code $\cC$ of $\Sigma^{N}$. For each codeword $\vx \in \cC$, let
\[ \cS(\vx) \eqdef \set*{\vx_{0+[n]}, \vx_{n-\Lo+[n]}, \vx_{2(n-\Lo)+[n]},  \ldots,\vx_{(k-1)(n-\Lo)+[n]}   }  \in \cX_{n,k}.\]
The code we construct is $\cD$, defined as,
\[ \cD \eqdef \set*{ \cS(\vx) ~:~ \vx \in \cC} \subseteq \cX_{n,k}. \]
\end{construction}

\begin{lemma}Let $\Lm>\Lo$. Then the code $\cD$ from Construction~\ref{Construction3} is a multi-strand  $(\Lm,\Lo,e)$-trace reconstruction code of $\cX_{n,k}$.
\end{lemma}

\begin{IEEEproof}
It is easy to see that an $(\Lm,\Lo,e)$-erroneous trace $\cY$ of $\cS(\vx)$ is also  an $(\Lm,\Lo,e)$-erroneous trace of $\vx$. Since $\cC$ is a trace reconstruction code, then for each $\vy \in \cY$,  we can determine its location in $\vx$. Hence, we can determine the index $i$ such that $\vy \in \cY_i$ and determine the  location of $\vy$ in $\vx_i$.
\end{IEEEproof}

\begin{lemma}[{{\cite[Lemma 16]{YehBarMarYaa:2023}}}] $\log \abs{\cX_{n,k}}=k(n-\log(k/\ee))+o(k)$ \footnote{We use $\ee$ to denote $\exp(1)$ in order to avoid confusion with $e$ which denotes the number of errors.}.
\end{lemma}

\begin{theorem}\label{thm:multilargeLo}
Suppose that $\limsup_{n\to \infty} \log k / n <1$, $\Lo = \ceil{\log (nk)} +(24e+13)\ceil{\log \ceil{\log (nk)}} + (4e+1)\ceil{\log (4e+1)} +20e+5$ and $\Lm > \Lo$. For sufficiently large $n$,  there is a multi-strand  $(\Lm,\Lo,e)$-trace reconstruction code of $\cX_{n,k}$ whose rate is $1-o(1)$.
\end{theorem}

\begin{IEEEproof} Let $N= k(n-\Lo)+\Lo$. Then  $\Lo \geq  \ceil{\log N} +(24e+13)\ceil{\log \ceil{\log N}} + (4e+1)\ceil{\log (4e+1)} +20e+5$. According to Corollary~\ref{cor:genrecon-1}, there is an $(\Lm,\Lo,e)$-trace reconstruction code $\cC$ of $\Sigma^n$ whose rate is $1-o(1)$. Applying Construction~\ref{Construction3} with this code, we obtain a  multi-strand  $(\Lm,\Lo,e)$-trace reconstruction code $\cD$ of $\cX_{n,k}$ with $\abs{\cD}=\abs{\cC}$. Note that 
\begin{align*}
\frac{N}{\log \abs{\cX_{n,k}}}&=\frac{k(n-\Lo)+\Lo}{k(n-\log(k/\ee))+o(k)} =\frac{n-\Lo+\Lo/k}{n-\log k +O(1) }\\
& = 1-\frac{\Lo-\log k -\Lo/k+O(1) }{n-\log k +O(1)}=1-O\parenv*{\frac{\log n}{n}}.
\end{align*}
Hence, the code rate is
\begin{align*}
R(\cD) =\frac{\log \abs{\cD}}{\log \abs{\cX_{n,k}}}=\frac{\log \abs{\cC}}{N} \frac{N}{\log \abs{\cX_{n,k}} }=(1-o(1))\parenv*{ 1-O\parenv*{\frac{\log n}{n}} }=1-o(1).
\end{align*}
\end{IEEEproof}

Now, we consider the case of $\Lo\leq \log (n k)$.
Assume that $\Lm= a\log (nk)$ and $\Lo=\gamma \Lm$ where $a>1$ and $0\leq a \gamma \leq 1$. Let
\[ L^* \eqdef (n-\Lo) \bmod (\Lm-\Lo). \]
We first present some upper bounds on the rate of multi-strand $(\Lm,\Lo)$-trace reconstruction codes. 

\begin{lemma}[{{\cite[In the proof of Lemma~8]{YehBarMarYaa:2023}}}]
\label{lm:combinnumeq}
For all $v\geq u\geq 0$,  $\log \binom{u+v}{u} < u(2\log \ee + \log v -\log u).$
\end{lemma}

\begin{lemma} \label{lemma:multismallLoUpBnd}
Suppose that  $\Lm= \ceil{a\log (nk)}$ and $\Lo=\ceil{\gamma \Lm}$ where $a>1$ and $0\leq a \gamma \leq 1$.
Let $\cC$ be a multi-strand $(\Lm,\Lo)$-trace reconstruction code of $\cX_{n,k}$. Then  it holds that
\[\frac{\log \abs{\cC}}{nk} \leq \parenv*{\frac{1-1/a}{1-\gamma}} \parenv*{1-\gamma\frac{\Lm}{n}}  +\frac{1/a-\gamma}{1-\gamma}\cdot\frac{L^*}{n}+O\parenv*{\frac{\log n}{n}}.\]
In particular, if $\log k = o(n)$, then the code rate satisfies
\[R(\cC) \leq \frac{1-1/a}{1-\gamma} +o(1), \]
and if $\log k =\kappa n$ where $0< \kappa <1$ is a real constant, then the code rate satisfies
\[ R(\cC) \leq \frac{1-a\gamma \kappa}{ 1-\kappa} \parenv*{\frac{1-1/a}{1-\gamma}}+\frac{1/a-\gamma}{(1-\gamma)(1-\kappa)} \cdot\frac{L^*}{n} +O\parenv*{\frac{\log n}{n}}. \] 
\end{lemma}

\begin{IEEEproof}For a sequence $\vx \in \Sigma^n$, let 
\[\hcY(\vx)\eqdef \set*{\vx_{i(\Lm-\Lo) +[\Lm]} ~:~ i\in \left[\frac{n-\Lo-L^*}{\Lm-\Lo}-1\right] } \cup \set{\vx[n-\Lm-L^*,n-1]}. \]
For a codeword $\cS=\set{\vx_0,\vx_1,\ldots,\vx_{k-1}}\in\cC$, let 
$\hcY(\cS) \eqdef \bigcup_{i=0}^{k-1} \hcY(\vx_i).$
Then $\hcY(\cS)$ is an $(\Lm,\Lo)$-trace of $\cS$. 

Since $\cC$ is an $(\Lm,\Lo)$-trace reconstruction code,  necessarily $\hcY(\cS)\neq \hcY(\cS')$ for any two different codewords $\cS$ and $\cS'$. It follows that  
\[\abs{\cC} \leq \abs*{\set*{\hcY(\cS) ~:~ \cS\in \cC }}.\]
Note that $\hcY(\cS)$ is a multiset consisting of $k\frac{n-\Lm-L^*}{\Lm-\Lo}$ sequences of $\Sigma^{\Lm}$ and $k$ sequences of $\Sigma^{\Lm+L^*}$. Hence, 
\begin{equation}\label{eq:codesize}
\abs{\cC} \leq \binom{k\parenv*{\frac{n-\Lm-L^*}{\Lm-\Lo}}+2^{\Lm}-1}{2^{\Lm}-1 } \cdot  \binom{k+2^{\Lm+L^*}-1}{2^{\Lm+L^*} -1}. 
\end{equation}
We denote the first binomial coefficient in \eqref{eq:codesize} as $A$ and the second one as $B$. Since $2^{\Lm} \geq (nk)^a > \frac{k(n-\Lm)}{\Lm-\Lo}$ and $2^{\Lm+L^*} >k$, according to Lemma~\ref{lm:combinnumeq}, we have that
\begin{align}
\frac{\log A}{nk} & < \frac{k}{nk} \parenv*{\frac{n-\Lm-L^*}{\Lm-\Lo}} \parenv*{2\log \ee +\Lm - \log(k(n-\Lm-L^*)) + \log (\Lm-\Lo)} \notag \\ 
 & = \frac{1-(\Lm+L^*)/n}{\Lm-\Lo}(\Lm-\log (nk) +O(\log \log (nk))) \notag  \\ 
 &= \parenv*{1-\frac{\Lm+L^*}{n}}\frac{\Lm-\log (nk)}{\Lm-\Lo} +O\parenv*{\frac{\log \log (nk)}{ \log (nk)}} \notag \\
 &= \frac{1-1/a}{1-\gamma} \parenv*{1-\frac{\Lm+L^*}{n}} +O\parenv*{\frac{\log \log (nk)}{ \log (nk)}}, \label{eq:logA}
\end{align}
and
\begin{align}
\frac{\log B}{nk} & < \frac{1}{n}\parenv*{2\log \ee +\Lm+L^*-\log k  }= \frac{(1-1/a)\Lm}{n} +\frac{L^*}{n}+O\parenv*{\frac{\log n}{n}}. \label{eq:logB}
\end{align}
Combining \eqref{eq:codesize}, \eqref{eq:logA} and \eqref{eq:logB}, we have that 
\begin{equation}\label{eq:coderate}
\frac{\log \abs{\cC}}{nk} \leq \parenv*{\frac{1-1/a}{1-\gamma}} \parenv*{1-\gamma\frac{\Lm}{n}} +\frac{1/a-\gamma}{1-\gamma}\cdot\frac{L^*}{n}+O\parenv*{\frac{\log n}{n}}.  
\end{equation}

If $\log k=o(n)$, then $\Lm / n = a \log(nk)/n=o(1)$ and $L^*/n<\Lm/n = o(1)$. It follows that
\[\frac{\log \abs{\cC}}{nk} \leq \parenv*{\frac{1-1/a}{1-\gamma}} \parenv*{1-o(1)} +o(1)=\frac{1-1/a}{1-\gamma}+o(1). \]
Recall that $\log \abs{\cX_{n,k}}=k(n-\log(k/\ee))+o(k)$. Hence, the code rate
\begin{align*}
R(\cC) = \frac{\log \abs{\cC}}{\log \abs{\cX_{n,k}}}= \frac{\log \abs{\cC}}{nk} \cdot\frac{nk}{k(n-\log(k/\ee))+o(k)}\leq \parenv*{\frac{1-1/a}{1-\gamma} +o(1)}\frac{1}{1-o(1)}=\frac{1-1/a}{1-\gamma} +o(1).
\end{align*}

If $\log k=\kappa n$ where $0<\kappa <1$ is a real constant, then 
\begin{align*}
\frac{nk}{\log \abs{\cX_{n,k}}}=\frac{nk}{ k(n-\log(k/\ee))+o(k) } =\frac{1}{1-\kappa +O(1/n)}=\frac{1}{1-\kappa}-O\parenv*{\frac{1}{n}}.
\end{align*}
Therefore, it follows from \eqref{eq:coderate} that the code rate satisfies
\begin{align*}
R(\cC) & = \frac{\log \abs{\cC}}{\log \abs{\cX_{n,k}}}= \frac{\log \abs{\cC}}{nk} \frac{nk}{ \log \abs{\cX_{n,k}}  } \\
& \leq \parenv*{\parenv*{\frac{1-1/a}{1-\gamma}} \parenv*{1-a\gamma\kappa} +\frac{1/a-\gamma}{1-\gamma}\frac{L^*}{n}+O\parenv*{\frac{\log n}{n}}} \parenv*{  \frac{1}{1-\kappa}-O\parenv*{\frac{1}{n}} }\\
& = \frac{1-a\gamma \kappa}{ 1-\kappa} \parenv*{\frac{1-1/a}{1-\gamma}} + \frac{1/a-\gamma}{(1-\gamma)(1-\kappa)} \cdot\frac{L^*}{n} +O\parenv*{\frac{\log n}{n}}.
\end{align*}
\end{IEEEproof}

\begin{corollary}\label{cor:vanishinglowbnd-1}
Suppose that $\log k =o(n)$. Let $\cC$ be a multi-strand $(\Lm,\Lo)$-trace reconstruction code of $\cX_{n,k}$. If $\Lm\leq \log (nk)+o(\log (nk))$, then  $R(\cC) =o(1)$.
\end{corollary}

\begin{IEEEproof}
Since $\cC$ is also a multi-strand $(\ceil{a\log(nk)},0)$-trace reconstruction code for any $a>1$, it follows from Lemma~\ref{lemma:multismallLoUpBnd}  that $R(\cC) \leq 1-1/a+o(1)$ for all $a>1$. Hence, $R(\cC)=o(1)$. 
\end{IEEEproof}

\begin{lemma}\label{lm:vanishinglowbnd-2}
Suppose that $\log k \leq\kappa n$ where $\kappa <1$ is a constant.  Let $\cC$ be a multi-strand $(\Lm,\Lo)$-trace reconstruction code of $\cX_{n,k}$. If $\Lm= \ceil{a\log (nk)}$ for some $a<1$, then  $R(\cC) =o(1)$.
\end{lemma}

\begin{IEEEproof}
The proof is similar to that of Lemma~\ref{lemma:multismallLoUpBnd}. In this case, we denote 
\[ \hcY(\vx)\eqdef \set{\vx_{0+[\Lm]},\vx_{1+[\Lm]},\ldots,\vx_{n-\Lm+[\Lm]}}.\]
Then  each $\hcY(\cS) = \bigcup_{i=0}^{k-1} \hcY(\vx_i)$ is still an $(\Lm,\Lo)$-trace, and it consists of $k(n-\Lm+1))$ sequences of $\Sigma^{\Lm}$, and so,
\[\abs{\cC} \leq \binom{k(n-\Lm+1) +2^{\Lm}-1}{2^{\Lm}-1}.\]
We observe that $k(n-\Lm+1) \geq  k(n-a\log n -a \log k)  \geq k\parenv*{ (1-a\kappa)n-a\log n} \geq  c nk$ for some constant $c$ and $2^{\Lm} \leq 2(nk)^a$. Since $a<1$, when $n$ is sufficiently large, we have that $k(n-\Lm+1) \geq 2^{\Lm}$. Using the inequality in Lemma~\ref{lm:combinnumeq}, we get that
\begin{equation}\label{eq:msbnd}
\frac{1}{nk}\log \binom{k(n-\Lm+1) +2^{\Lm}-1}{2^{\Lm}-1} \leq\frac{2^{\Lm}}{nk} \parenv*{ 2\log \ee + \log (k(n-\Lm+1)) -\Lm}.
\end{equation}
Noting that $kn > k(n-\Lm+1) \geq  c nk$, we have that $\log ( k(n-\Lm+1))=\log( nk )-O(1)$.
Continuing~\eqref{eq:msbnd}, 
\begin{align*}
\frac{1}{nk}\log \binom{k(n-\Lm+1) +2^{\Lm}-1}{2^{\Lm}-1} &\leq\frac{2^{\Lm}}{nk} \parenv*{ 2\log \ee + \log (k(n-\Lm+1)) -\Lm}\\
&\leq\frac{2^{\Lm}}{nk} \parenv*{(1-a)\log(nk)+O(1)}\\
&=\frac{(1-a)\log (nk)+O(1)}{(nk)^{1-a}}=o(1).
\end{align*}
Hence,
\[R(\cC)=\frac{\log \abs{\cC}}{\log\abs{\cX_{n,k}}}=\frac{\log \abs{\cC}}{nk}\cdot\frac{nk}{\log\abs{\cX_{n,k}}}    = o(1).\]
\end{IEEEproof}

\begin{lemma}\label{lm:vanishinglowbnd-3}Suppose that $k\leq 2^n$.
 Let $\cC$ be a multi-strand $(\Lm,\Lo)$-trace reconstruction code of $\cX_{n,k}$. If $\Lm \leq \log (nk)+o(\log (nk))$ and $\Lm-\Lo=\Theta(\log (nk))$, then  $R(\cC) =o(1)$.
\end{lemma}

\begin{IEEEproof}It suffices to consider the case of $\Lm=\log (nk)+o(\log(nk))$. In this case, we denote 
\[\hcY(\vx)\eqdef \set*{\vx_{i(\Lm-\Lo) +[\Lm]} ~:~ i\in \left[\frac{n-\Lo-L^*}{\Lm-\Lo}\right] } \cup \set{\vx[n-\Lm,n-1]}.\]
Since $\Lm-L^*\geq \Lo$, each $\hcY(\cS) = \bigcup_{i=0}^{k-1} \hcY(\vx_i)$ is still an $(\Lm,\Lo)$-trace, and it consists of $k\parenv*{\frac{n-\Lo-L^*}{\Lm-\Lo}+1}$ sequences of $\Sigma^{\Lm}$. Hence, we have that 
\[\abs{\cC} \leq \binom{\frac{k(n+\Lm-2\Lo-L^*)}{\Lm-\Lo} +2^{\Lm}-1}{2^{\Lm}-1}.\]
Since  $\Lm=\log(nk)+o(\log (nk))$, we have $\frac{k(n+\Lm-2\Lo-L^*)}{\Lm-\Lo}<2^{\Lm}$ . Using the inequality in Lemma~\ref{lm:combinnumeq}, we get that
\begin{align}
&\frac{1}{nk}\log \binom{\frac{k(n+\Lm-2\Lo-L^*)}{\Lm-\Lo}  +2^{\Lm}-1}{2^{\Lm}-1} \notag\\
\leq & \frac{k(n+\Lm-2\Lo-L^*) }{(\Lm-\Lo)nk}\parenv*{2\log \ee + \Lm - \log(k(n+\Lm-2\Lo-L^*))+\log(\Lm-\Lo)  }\notag\\
\leq & \frac{k(n+\Lm-2\Lo-L^*) }{(\Lm-\Lo)nk}\parenv*{2\log \ee + \Lm - \log(k(n-\Lo))+\log(\Lm-\Lo) \label{eq:msbnd-2}  }.
\end{align}
Since $\Lm\leq \log (nk)+o(\log(nk))$ and $\Lm-\Lo=\Theta(\log(nk) )$, we have that $\Lo\leq c_1\log (nk)\leq c_2 n$ for some constants $c_1,c_2<1$. It follows that $\log(k(n-\Lo)) = \log (nk)-O(1)$.
Continuing \eqref{eq:msbnd-2}, we have that 
\begin{align*}
&\frac{1}{nk}\log \binom{\frac{k(n+\Lm-2\Lo-L^*)}{\Lm-\Lo}  +2^{\Lm}-1}{2^{\Lm}-1} \\
\leq & \frac{k(n+\Lm-2\Lo-L^*) }{(\Lm-\Lo)nk}\parenv*{2\log \ee + \Lm - \log(k(n-\Lo))+\log(\Lm-\Lo)  }\\
\leq & \frac{k(n+\Lm-2\Lo-L^*) }{(\Lm-\Lo)nk}\parenv*{2\log \ee + \Lm - \log (nk) +O(\log \log (nk)) }\\
= & \parenv*{1+\frac{\Lm-2\Lo-L^*}{n}}\frac{o(\log (nk)) }{\Lm-\Lo}=o(1),
\end{align*}
where the last equality holds since $\Lm-\Lo=\Theta(\log (nk))$.
Hence,
\[R(\cC)=\frac{\log \abs{\cC}}{\log\abs{\cX_{n,k}}}=\frac{\log \abs{\cC}}{nk}\cdot\frac{nk}{\log\abs{\cX_{n,k}}}    = o(1).\]
\end{IEEEproof}

\begin{remark}
We note that the condition $\Lm-\Lo=\Theta(\log(nk))$ in Lemma~\ref{lm:vanishinglowbnd-3} cannot be removed. A counterexample is the $(\Lm,\Lo,e)$-trace reconstruction codes of rate $1-o(1)$ in Theorem~\ref{thm:multilargeLo}, where $\Lo = \ceil{\log nk} +(24e+13)\ceil{\log \ceil{\log nk}} + (4e+1)\ceil{\log (4e+1)} +20e+5$ and $\Lm \geq \Lo +1$.
\end{remark}

Note that a multistrand $(\Lm,\Lo,e)$-trace reconstruction code is also a  multistrand $(\Lm,\Lo)$-trace reconstruction code. Hence, the upper bounds in Lemmas~\ref{lemma:multismallLoUpBnd}--\ref{lm:vanishinglowbnd-3}  also work for multistrand $(\Lm,\Lo,e)$-trace reconstruction codes.

In the following, we study the lower bounds. 

\begin{theorem}
\label{thm:multismallLoLowBnd}
Let $\Lm= \ceil{a\log (nk)}$ and $\Lo=\ceil{\gamma \Lm}$, where $a>1$ and $0\leq a\gamma \leq 1$. For all sufficiently large $n$, 
\begin{enumerate}
\item if  $\log k=o(n)$, then there is a multi-stand $(\Lm,\Lo,e)$-trace reconstruction code $\cD$ of $\cX_{n,k}$ of rate
\[R(\cD) =  \frac{1-1/a}{1-\gamma} -o(1);\]
\item if $\log k =\kappa n$ where $0< \kappa <1$ is a real constant, then there is a multi-strand $(\Lm,\Lo,e)$-trace reconstruction code $\cD$ of $\cX_{n,k}$ of rate
\[ R(\cD) =  \frac{1-a\gamma \kappa}{ 1-\kappa} \parenv*{\frac{1-1/a}{1-\gamma}} -o(1). \]
\end{enumerate}
\end{theorem}

\begin{IEEEproof} Let $N= k(n-\Lo)+\Lo$. Then  $\Lm \geq \ceil{a \log N}$.  According to Theorem~\ref{thm:tracerecon}, there is an $(\Lm,\Lo,e)$-trace reconstruction code $\cC$ of $\Sigma^N$ whose rate is $\frac{1-1/a}{1-\gamma}-o(1)$. Applying Construction~\ref{Construction3} with this code, we obtain a  multi-strand  $(\Lm,\Lo,e)$-trace reconstruction code $\cD$ of $\cX_{n,k}$ with $\abs{\cD}=\abs{\cC}$. Note that
\begin{align*}
\frac{N}{\log \abs{\cX_{n,k}}}&=\frac{k(n-\Lo)+\Lo}{k(n-\log(k/e))+o(k)} =\frac{n-\Lo+\Lo/k}{n-\log k +O(1) }\\
& = 1-\frac{\Lo-\log k -\Lo/k+O(1) }{n-\log k +O(1)}=1-\frac{(a\gamma-1)\log k +O(\log n)}{n-\log k +o(1)}.
\end{align*}

If  $\log k=o(n)$, then ${N}/{\log \abs{\cX_{n,k}}}=1-o(1)$, and so, we have that 
\begin{align*}
R(\cD) =\parenv*{ \frac{1-1/a}{1-\gamma}-o(1) } (1-o(1)) =\frac{1-1/a}{1-\gamma}-o(1).
\end{align*}
If  $\log k=\kappa n$, then 
\[\frac{N}{\log \abs{\cX_{n,k}}}=1-\frac{(a\gamma -1)\kappa}{1-\kappa}-o(1)=\frac{1-a\gamma \kappa}{1-\kappa}-o(1),\] 
and so, we have that
\begin{align*}
R(\cD) =\parenv*{ \frac{1-1/a}{1-\gamma}-o(1) } \parenv*{\frac{1-a\gamma \kappa}{1-\kappa}-o(1)} =\frac{1-a\gamma \kappa}{ 1-\kappa} \parenv*{\frac{1-1/a}{1-\gamma}} -o(1).
\end{align*}
\end{IEEEproof}

When $\log k =o(n)$ or when $\log k =\kappa n$ and $L^* =o(n)$, the lower bounds in Theorem~\ref{thm:multismallLoLowBnd} asymptotically achieve the upper bound in Lemma~\ref{lemma:multismallLoUpBnd}.

Next, we  show that when $\log k =\kappa n$ and $\Lo=0$, if $L^* \leq \Lm-(1+\epsilon)\log (nk)= (a-1-\epsilon) \log (nk)$ for a  positive $\epsilon$ which is independent of $n$, then the upper bound in  Lemma~\ref{lemma:multismallLoUpBnd} still can be achieved.

\begin{construction}
\label{Construction4}
Suppose that $\Lm=\ceil{a \log (nk)}$ and $\Lo=0$. Denote $\ban\eqdef \frac{n-L^*}{\Lm}$ and $K\eqdef \ceil{\sqrt{\log (nk)}}$. Let $I\eqdef \ceil{\log (\ban k)}$ and $r_I\eqdef \ceil{(3d+8)\log I}$ where $d=2e+1$. Then according to Theorem~\ref{thm:singleSDS}, there is a collection of $(3\ceil{\frac{3}{2}\log (I+r_I)}+\ell,d)$-WWL sequences $\vc_0,\vc_1,\ldots,\vc_{2^I-1}\in \Sigma^{I+r_I}$  such that the concatenation $\vc_0\circ \vc_1\circ \cdots \circ \vc_{2^I-1}$ is an $(I+r_I,d)$-SD sequence.
 
Denote $n'\eqdef  \ban(\Lm-(I+r_I+K+\ell))+L^*$. Let $\cE_{\mathtt{WWL}}$ be the encoder in \cite[Algorithm~2]{LevYaa:2017} which can encode sequences of $\Sigma^{n'-d}$ into $(\ceil{K/4},d)$-WWL sequences\footnote{Note that $n'=\Theta(n)$ and $K=\sqrt{\log(nk)}=\Theta(\sqrt{n})$. Hence, $K/4 \gg \cF(n',d)= \log n' + (d-1)\log \log n'+O(1)$. Then according to Lemma 19 in \cite{LevYaa:2017}, the encoder $\cE_{\mathtt{WWL}}$ does work. }  of $\Sigma^{n'}$. For a message $\vm=\vm_0\circ \vm_1\circ \cdots \circ \vm_{k-1}$ where $\vm_i\in \Sigma^{n'-d}$ for $i \in[k]$,  let $\vv_i\eqdef \cE_{\mathtt{WWL}}(\vm_i) $ for all $i \in [k]$. We partition each $\vv_i$ into $\ban+1$ substrings as follows:
\[\vv_i=\vv_{i,0}\circ \vv_{i,1} \circ \cdots \vv_{i,\ban-1} \circ \vv_{i,\ban}\]
where $\abs{\vv_{i,j}}= \Lm-(I+r_I+K+\ell)$ for $j\in [\ban]$ and $\abs{\vv_{i,\ban}}=L^*.$ 

Denote $\vp \eqdef 0^K \circ \vu$ where $\vu$ is a $d$-auto-cyclic sequence of length $\ell$. For each $i \in [k]$, let 
\[\vw_i = \vv_{i,0} \circ \vp  \circ \vc_{i\ban} \circ  \vv_{i,1} \circ \vp \circ \vc_{i \ban+1} \circ \cdots \circ  \vv_{i,\ban-1} \circ \vp \circ \vc_{(i+1)\ban-1} \circ \vv_{i,\ban}.  \]

Output $\set{\vw_0,\vw_1,\ldots,\vw_{k-1}}$ as the codeword which encodes the message  $\set{\vm_0,\vm_1,\ldots,\vm_{k-1}}$. The image of the mapping described here is the constructed code.
\end{construction}

\begin{lemma}
Suppose that $L^* \leq \Lm-(1+\epsilon)\log (nk)$ for a positive $\epsilon$ which is independent of $n$. Then the code obtained in Construction~\ref{Construction4} is a multi-strand $(\Lm,0,e)$-trace reconstruction code of $\cX_{n,k}$.
\end{lemma}

\begin{IEEEproof}[Sketch of proof]  Let  $\vy$  be a length-$\Lm$ substring of $\vw_i$ for some $\vw_i \in \set{\vw_0,\vw_1,\ldots,\vw_{k-1}}$. Note that $L^* \leq \Lm-(1+\epsilon)\log (nk)$ and $\abs{\vp\circ \vc_{j}}=K+\ell+I+r_I <(1+\epsilon)\log (nk)$. Then $\vy$ must contain either a copy of $\vp\circ \vc_{i\ban+j}$ or a suffix of $\vp\circ \vc_{i\ban+j}$ together with a prefix of $\vp\circ \vc_{i\ban+j+1}$. Since  $\vv_i$'s and $\vc_j$'s are WWL sequence, even if $\vy$ suffers from $e$ errors, we can still  locate their position in $\vy$ by searching for the marker $\vp$. Then we can run the locating algorithm of the robust positioning sequence  $\vc_0\circ \vc_1\circ \cdots \circ \vc_{2^I-1}$  to determine the index $i\ban+j$ or $i\ban+j+1$, and hence the location of $\vy$.
\end{IEEEproof}

\begin{theorem}\label{thm:multinullLoLowBnd}
 Suppose that $\log k=\kappa n$, $\Lm=\ceil{a \log (nk)}$ and $\Lo=0$, where $0<\kappa <1$ and $a >1$.  If  $L^* \leq \Lm-(1+\epsilon)\log (nk)$ for a fixed positive $\epsilon$ which is independent of $n$, then there is a multi-strand $(\Lm,0,e)$-trace reconstruction code which has code rate
 \[  \frac{1-1/a}{1-\kappa}+ \frac{1}{a(1-\kappa) } \cdot\frac{L^*}{n} -o(1) \]
\end{theorem}
\begin{IEEEproof}Note that 
\begin{align*}
\frac{n'-d}{n} &= \frac{\ban(\Lm-(I+r_I+K+\ell))+L^*-d}{n}= \frac{n-\ban(I+r_I+K+\ell)-d}{n} \\
&= 1-\frac{1-L^*/n}{\Lm}(I+r_I+K+\ell)-O\parenv*{\frac{1}{n}}\\
&= 1-\parenv*{1-\frac{L^*}{n}} \frac{\log (nk)+O(\sqrt{\log (nk)})}{a\log (nk)}-O\parenv*{\frac{1}{n}}\\
&= 1-\frac{1}{a}+\frac{L^*}{an}-O\parenv*{\frac{1}{\sqrt{\log (nk)}}}.
\end{align*}
Hence, the code rate is
\[\frac{ (n'-d)k }{\log  \abs{\cX_{n,k}}} =  \frac{(n'-d)k}{nk} \frac{nk}{\log  \abs{\cX_{n,k}}}= \parenv*{1-\frac{1}{a}+\frac{L^*}{an}-o(1)}\parenv*{\frac{1}{1-\kappa}-o(1)}
=\frac{1-1/a}{1-\kappa}+ \frac{1}{a(1-\kappa) } \frac{L^*}{n} -o(1). \]
\end{IEEEproof}

Finally, we note that the 
multi-strand $(\Lm,0,e)$-trace reconstruction code in Construction~\ref{Construction4} only guarantees recovering message  from reliable $(\Lm,0,e)$-erroneous traces, the occurrence of which might be rare since $\Lo=0$ and each symbol is usually included in a small number of substrings in $\cY$. Nevertheless,  
we can use a $(k,2^{(n'-d)(k-r_o)},2\tau+1)_{2^{n'-d}}$ code to encode the message, like what we have done in Construction~\ref{Construction2}, so that even if there are in total $\tau$ errors in $\cY$, we still can decode the message. The rate of this trace reconstruction code is \[\parenv*{1-
\frac{r_o}{k}}\parenv*{\frac{1-1/a}{1-\kappa}+ \frac{1}{a(1-\kappa) } \cdot\frac{L^*}{n} }-o(1).\]

\bibliographystyle{IEEEtranS}
\bibliography{reference}
\end{document}